\documentclass{aastex}

\usepackage{longtable}
\usepackage{subfigure}

\begin{document}

\title{Lithium Abundances of the Super-Metal-Rich\\ Open Cluster NGC 6253\footnote{Publication
number 50 of the WIYN Open Cluster Study}}
\author{Jeffrey D. Cummings}
\affil{Departamento de Astronom\'ia, Universidad de Concepci\'on}
\affil{160-C Casilla, Concepci\'on, Chile}
\email{jcummings@astro-udec.cl}
\author{Constantine P. Deliyannis}
\affil{Department of Astronomy, Indiana University}
\affil{Bloomington, IN 47405-7105, USA}
\email{con@astro.indiana.edu}
\author{Barbara Anthony-Twarog, Bruce Twarog}
\affil{Department of Physics and Astronomy, University of Kansas}
\affil{Lawrence, KS 66045-7582, USA}
\email{bjat@ku.edu, btwarog@ku.edu}
\and
\author{Ryan M. Maderak}
\affil{Department of Astronomy, Indiana University}
\affil{Bloomington, IN 47405-7105, USA}
\email{maderak@astro.indiana.edu}

\begin{abstract}
High-resolution CTIO 4-m/HYDRA spectroscopy of the super-metal-rich open cluster NGC 6253
([Fe/H]=+0.43$\pm$0.01) has been used to study the stellar lithium (Li) abundances near the 
cluster's turnoff.  NGC 6253 greatly expands the range of [Fe/H] for clusters that have a Li
abundance analysis.  This
is important for studying the complicated effects of, and potential correlations with, stellar Fe
abundance on surface Li abundance.  Comparisons to the younger and less-metal-rich Hyades and to
the similarly-aged but solar-metallicity M67 show that NGC 6253's Li abundances are qualitatively
consistent with the prediction, from Standard Stellar Evolution Theory, that higher-metallicity
stars have a greater Li depletion.  Comparison with M67 provides evidence that the more-metal-rich
NGC 6253 had a higher initial Li, which is consistent with expectations from models
of Galactic Li production.  NGC 6253 is also compared to the intermediate-aged NGC 3680, NGC 752,
and IC 4651 open clusters.  Comparison of the Li-gap positions in all six clusters shows: a) the 
gap's position in T$_{eff}$ is independent of metallicity, but b) higher-metallicity clusters have their
gaps in higher-mass stars.  In addition, the Li gap's position is shown not to evolve with age, 
which provides an important constraint for the non-standard depletion mechanisms that may create 
the Li gap.
\end{abstract}


\section{Introduction}

The study of Li in open clusters provides invaluable information about physical processes
occurring in the interior of stars.  This is because in stellar envelopes, energetic protons will
break apart the fragile Li nucleus at T $\ge$ 2.5 million K.  Observations of dwarfs across a
broad range of star
clusters have shown that surface Li abundances are depleted in most stars (Deliyannis 2000;
Jeffries 2000; Sestito \& Randich 2005).  The standard model (no rotation, no mass loss, no magnetic
fields, and no diffusion) for this surface Li depletion in Population I dwarfs (Deliyannis
et al. 1990; Pinsonneault 1997, hereafter P97) states that when a dwarf is predominantly
convective during its pre-main
sequence (PMS) phase, the material at the stellar surface reaches deep into the interior and
gradually depletes its Li abundance.  Higher-mass A and F dwarfs ($\ge$ 1.3 M$_{\odot}$) remain
in the PMS phase only briefly, so only a negligible amount of surface Li depletion occurs before
they reach the main sequence.  Lower-mass dwarfs remain in the PMS phase longer and have a higher
temperature (and density) at the base of their surface convection zone (SCZ), so they deplete a
greater amount of surface Li.  Subsequently, on the main sequence only the late G, K, and M dwarfs
have SCZs that are deep enough to reach regions with the necessary conditions for Li destruction,
and over time they continue to gradually deplete their surface Li abundance.  The standard model
also predicts that Li depletion has a strong dependence on metallicity, where the more-metal-rich
stars deplete surface Li at a faster rate.  This is because a higher metallicity greatly increases
the opacity in the outer layers of stars, and an increase in opacity increases the depth of the
SCZ.  The standard model of Li depletion assumes that this Li depletion through convective
processes, predominantly during the PMS, is the sole Li-depletion mechanism.

Open cluster observations consistently show, however, that the standard model predicts insufficient Li
depletion in the majority of, if not all, observed stellar temperature ranges (P97).  Especially
surprising are the severe Li depletions (up to at least $\sim$2 dex) observed in mid-F cluster dwarfs
(the ``Li gap"), which are expected to have \textit{negligible} standard PMS Li depletion.  This Li
gap was first discovered in the $\sim$650 Myr-old Hyades (Boesgaard \& Tripicco 1986), followed by
its discovery in the the older (1.45$\pm$0.1 Gyr; Anthony-Twarog et al. 2009; hereafter AT09) NGC
752 (Hobbs \& Pilachowski 1986) and in the
Hyades-aged Praesepe ($\sim$650 Myr; Soderblom et al. 1993a).  Li data of F dwarfs in the 100 Myr-old
Pleiades (Boesgaard et al. 1988) confirmed the standard prediction of little or no PMS depletion
in F dwarfs, which implies that some unknown mechanism(s) acting during the \textit{main sequence}
has(have) created the Li gap in the older clusters.  
Pleiades data for the cooler G and K dwarfs (Soderblom et al. 1993b), stars that have
progressively deeper SCZs, also confirmed the overall standard prediction for Li depletion as a
function of T$_{eff}$ in these stars.  However, in contrast to standard theory, G dwarfs continue to
deplete Li during the main sequence, as evidenced by data in older clusters (e.g. see reviews by
Jeffries 2000; Deliyannis 2000), suggesting that they, too, are affected by additional Li-depleting
mechanisms.  A variety of mechanisms have been proposed to explain this additional Li depletion.  
For the Li gap, proposed mechanisms include diffusion (Richer \& Michaud 1993), mass loss (Schramm,
Steigman, \& Dearborn 1990), and gravitational-wave-induced mixing (Garcia Lopez \& Spruit 1991).  For
both the Li gap and cooler dwarfs, proposed mechanisms include rotationally-induced mixing 
(Pinsonneault et al. 1989; Pinsonneault et al. 1990), the combined effects of diffusion and 
rotationally-induced mixing (Chaboyer et al. 1995a and 1995b), and the combined effects of diffusion, 
rotationally-induced mixing, and gravity waves (e.g. Charbonnel \& Talon 2005), where diffusion is
more important in F dwarfs, and gravity waves are more important in G dwarfs.

Studying this additional Li depletion and comparing it to models are challenging because determination
of the total absolute magnitude of the Li depletion requires knowledge of the \textit{initial} Li
of the cluster, i.e., the presumed uniform Li
abundance of a cluster before any Li depletion occurs.  But, to zeroth order, the initial Li of 
near-solar-metallicity open clusters should be similar to the meteoritic (initial solar) abundance
of 12 + log (N$_{Li}$/N$_H$) == A(Li) = 3.31$\pm$0.04 (Anders \& Grevesse 1989) and to what is observed
in very-young open clusters, which are generally in the range of A(Li)=3.0 to 3.4 (e.g. IC 2602, 
Randich et al. 1997; NGC 2264, King 1998; Blanco 1, Jeffries \& James 1999; Alpha Per, Balachandran
et al. 1996; Pleiades, Soderblom et al. 1993b; NGC 2516, Jeffries et al. 1998).
This meteoritic A(Li) results from a combination of Li produced during Big Bang Nucleosynthesis (BBN)
and Galactic Li production.  Regardless of whether the Big Bang Li abundance (Li$_{BBN}$) is the
value observed in the near-uniform plateau of Li abundances in halo dwarfs (2.09$\pm$0.19/0.13, Ryan
et al. 2000), or the value inferred from WMAP in the context of standard BBN (2.72$\pm$0.06; Cyburt 
et al. 2008; note that either scenario for Li$_{BBN}$ has issues), the implication is that
the significantly higher meteoritic Li abundance requires that the Galaxy has produced most of the
Li that is around today.  Additionally, there is evidence from observations of field
dwarfs that there is a positive correlation between a star's [Fe/H] and its initial Li (Ryan et al.
2001; Travaglio et al. 2001).  Several mechanisms have been proposed that can produce Li in the
Galaxy: cosmic ray spallation of CNO and He; Type II Supernovae; the Be-7 transport mechanism in AGB
stars plus mass loss; and novae (Romano et al. 1999; Ryan et al. 2001; Travaglio et 
al. 2001).  These mechanisms, over time, will increase the Li abundance in the local interstellar 
medium and may correlate with local Fe production processes, which would be observed as an initial-Li
versus [Fe/H] correlation in stars.  However, despite considerable effort, Galactic chemical 
evolution models have not been able to match the trends in field-dwarf abundances or 
meteoritic Li; even in the best case, where all these mechanisms work in concert, the total Li
produced by the time Fe reaches its solar value falls below the meteoritic Li.  Perhaps the study of
cluster Li abundances across a broad range of [Fe/H] may be able to provide insight on the Galactic
production of Li, and an understanding of Galactic Li production may provide more precise initial
cluster Li abundances and, therefore, more precise absolute Li depletions.

With the strong dependence of Li depletion on [Fe/H] and the potential correlation of initial Li
versus [Fe/H], NGC 6253 provides a critically unique opportunity to examine the effects of
metallicity.  Evidence from photometry (e.g. Twarog 2003; hereafter T03) and spectroscopy (e.g.
Anthony-Twarog et al. 2010; hereafter AT10) suggests that NGC 6253 is one of the most-metal-rich
open clusters known, if not the most-metal-rich cluster known, with a spectroscopic metallicity
of 0.43$\pm$0.01 (AT10).  (See section 5.1 for a more detailed discussion.)  NGC 6253 is also an
older open cluster with an age of roughly 3 Gyr, allowing for an informative comparison with the
similarly-aged but solar-metallicity M67 (Pasquini et al. 2008; Jones et al. 1999).  Comparisons to
the significantly younger but well studied Hyades and to the intermediate-aged NGC 752, NGC 3680, 
and IC 4651 open clusters (AT09) are also of interest for analyzing the effects
of both age and metallicity.  Unfortunately, NGC 6253 cannot be compared directly to a standard 
Li-depletion model because its high [Fe/H] is beyond the range of published models.

In section 2, we discuss our NGC 6253 observations, data, and data-reduction techniques.  In
section 3, we discuss stellar radial velocities and cluster membership.  In section 4, we
discuss our adopted models and methods for deriving stellar parameters.  In section 5, we discuss
our Fe and Li analysis techniques.  In section 6, we discuss our Li abundance results for NGC 6253
and compare these to previous results from the Hyades, M67, NGC 3680, NGC 752, and IC 4651.  In
section 7, we discuss what these comparisons mean for a variety of models that may create
the Li gap in F dwarfs.  Lastly, we summarize our results in section 8.

\section{Data Observations and Reductions}

We obtained spectra of 54 candidate members in the turnoff region of NGC 6253 using the Blanco
4-meter telescope at the Cerro-Tololo-Inter-American Observatory with the HYDRA II multi-object
fiber-based spectrograph.  Observations of virtually identical HYDRA fiber configuration were
taken in 2005, 2006, and 2007.  Slight configuration changes were made over the three years 
because several fibers became unusable.  These three
observing runs were necessary because of poor observing conditions during both the 2005 and
2006 observations.  Improved conditions during the 2007 observations allowed for a majority of our
signal to be gathered during that time.  A second NGC 6253 configuration of blue stragglers
and giants was also observed.  Calibration images (dome flats, sky spectra, wavelength
calibration lamps, and etalons) were taken in large-circle configuration for the 2005 observations,
and we took all calibrations in the actual NGC 6253 HYDRA configuration during both 2006 and 2007.
Taking calibrations in configuration is ideal because it provides consistent fiber throughputs, 
which can change due to altering a fiber's position or magnitude of bending.  These same data were
used in the spectroscopic analysis of AT10.

Our cluster spectra cover a wavelength range from 6520 to 6800 \AA\, with a dispersion of 0.15
\AA\, per pixel.  The varying system throughput across this observed wavelength region means that
the spectrum's signal to noise (S/N) depends on wavelength.  However, unlike in
AT10, where a broad range of lines were of interest and several relatively line-free regions were
used to determine an appropriate overall S/N, in this paper we are focusing only on the Li
feature at 6708 \AA.  Therefore, all S/N measurements are based on the line-free region
near the Li feature at 6680 to 6695 \AA.  The S/N for all 54 observed stars range from 39 to 196
per pixel, with 30 stars having a S/N of at least 100 per pixel and only 4 stars having a S/N of 
less than 70 per pixel.

The steps to create our final spectra are briefly reviewed here (see also AT10).  We started with the
standard IRAF\footnote{IRAF is distributed by the National Optical Astronomy Observatory, which is operated
by the Association of Universities for Research in Astronomy, Inc., under cooperative agreement with
the National Science Foundation.} image processing steps, which are: fitting the overscan region and
applying the correction; subtracting the bias; and applying the fit and divided flat field to remove
the pixel-to-pixel variation.
At this stage the variation in throughput was not corrected with the flat field because we wanted
to retain the detailed information about the signal received.  The next steps were tracing the
individual fiber apertures with the dome flats, subtracting the scattered light, removing cosmic rays
with the L.A. Cosmic program for spectra (van Dokkum 2001), and extracting the individual spectra.  
During the 2005 and 2006 observations, contamination of the ThAr lamps required the use
of solar spectra for wavelength calibration of the night's series of etalons.  During the 2007 
observations, the ThAr calibration lamps were used.  Each night's series of etalons had at least
one etalon taken
immediately after the primary calibration spectrum (solar or ThAr), and this was followed by
additional etalons taken throughout and at the end of the night at roughly 2 hour intervals.  A
properly calibrated series of etalons each night gives a time-dependent wavelength calibration 
to account for possible wavelength shifts during the night.

In addition to the 54 fibers placed on stars in our configuration, we placed 30 fibers on sky
positions, which provided an excellent measure of the sky background.  Fibers vary in their total
throughput and in their wavelength-dependence of the throughput, so fiber-to-fiber throughput
corrections were determined using each day's high-signal daytime sky (solar spectrum) observations,
which provided a uniform and bright source of illumination of appropriate color to each fiber.  It
was assumed that the relative aperture throughput did not change significantly during the night.  
Application of this throughput correction to the entire night of cluster observations normalized the
multiple sky-background observations to the same scale in each exposure, in both sky fibers and 
object fibers.  In each exposure, the sky background in the object fibers was subtracted using the
median combination of all available sky fibers.  The next step was applying a Doppler correction 
to each object image based on the Earth's orbital velocity.  This correction provided a uniform 
radial-velocity zero point for all stellar spectra and allowed for precise radial-velocity 
measurements.  Finally, the individual object spectra were all co-added and continuum-fitted to
produce our final spectra.

\section{Radial Velocities, Binarity, and Membership}

The radial velocities, \textit{v sin i}, binarity information, and membership results in this paper 
have all been taken directly from AT10, where they have used the IRAF task \textit{fxcor}.    
Figure 1 shows a histogram of all measured radial velocities, and a majority of the observed 
stars create a narrow and approximately Gaussian radial-velocity distribution. To improve 
statistics, this histogram also includes the radial velocities of the giants and blue stragglers 
that were presented in AT10.  Single-star cluster
members should all have a radial velocity within the observed Gaussian distribution.  Stars that
are outside of this member distribution are either non-members or member binary systems; both cases
are removed from the cluster sample because determining spectroscopic abundances from binaries is
problematic.  This distribution and the cross-dispersion profiles from \textit{fxcor} imply that in
our observation of 54 turnoff candidates there are
41 radial-velocity members that show no evidence for binarity and 5 probable members that do show
evidence for binarity.  This high percentage of probable members demonstrates the strength of
photometric member determination when based on high-quality
photometry.  AT10 derive an average radial velocity for the single-star radial-velocity members of
-29.32$\pm$1.30 km/s.  The radial velocities for all probable single-star members in the turnoff
configuration are listed in Table 1.  Additionally, Maderak et al. (2012, in preparation; hereafter
M12) have
independently determined radial velocities of an identical dwarf sample with observations of the
oxygen-triplet region near 7770 \AA\, observed during June 2007 at the Blanco 4-meter with HYDRA.
M12 used the method of multiple
Doppler-shift measurements across the full spectral range by comparing observed line centers to
their rest wavelengths.  M12's independent data and radial velocity analysis found \textit{exactly}
the same sample of 41 radial-velocity turnoff members that were found in AT10.  Are all of these 
radial-velocity members single-star cluster members?  

Contamination in the Gaussian distribution by non-members should be minimal, but it cannot be
ruled out.  The number of potential non-members falling within the cluster's radial-velocity
distribution can be approximated from the distribution of non-members, which is partially shown
in Figure 1.  The average non-member distribution
estimates that there are 3 contaminating non-members in our Gaussian cluster distribution width of
6 km/s, a minor contribution in comparison to our 51 turnoff and giant radial-velocity members.
The proper-motion (PM) study of NGC 6253 by Montalto et al. (2009) has been used as a further
membership check with 27 of our 41 turnoff radial-velocity members having PM measurements.  Our
radial-velocity memberships and their PM memberships are mostly consistent.  Out of our 27
radial-velocity members that have Montalto et al. (2009) PM measurements, 2 have PM membership
probabilities of less than
5\%, another 2 have PM membership probabilities of roughly 50\%, and all others have PM membership
probabilities from 77\%\, to 96\%.   Considering both membership criteria for this paper, we do not 
consider the 2 radial-velocity members with a PM membership of less than 5\%\, to be members, but we
do consider the 2 radial-velocity members with a PM membership of roughly 50\%\, to be members.  
Having 2 of these 27 radial-velocity members be non-members is consistent with our statistical 
estimate of non-member contamination, as discussed above.  The published PM membership probabilities
are included in Table 1.  If additional
non-members remain in our sample of 39, we would expect them to have much lower [Fe/H], 
perhaps near-solar or even less.  However, not one star from either AT10's or M12's independent 
determinations of [Fe/H] lies below +0.24, and all are consistent with a distribution of [Fe/H] 
about the (super-metal-rich) cluster mean.  These findings provide additional evidence that all 39
dwarfs are cluster members.  We cannot absolutely rule out the possibility that a few member 
binaries (especially ones with extremely long periods) remain in the sample of 39, whose 
radial velocity just happens to coincide with the cluster mean at the time of 
observation; however, a) AT10 already threw out stars whose radial velocities varied between 2005,
2006, and 2007, and b) the 39 dwarfs show no evidence of binarity from \textit{fxcor}.  For these 
reasons, it is unlikely that any of the 39 dwarf members are binaries.  

\section{Stellar Parameters and Atmospheres}

Appropriate stellar atmospheres are needed to determine elemental abundances from equivalent-width
measurements or spectral synthesis.  All stellar parameters have been taken directly from AT10, 
where the full details of atmospheric parameter determinations are discussed.
For this study we have used Kurucz (1992) model atmospheres with no
convective overshoot.  Four stellar parameters are required to create model atmospheres: T$_{eff}$,
[Fe/H], surface gravity, and microturbulence.  The T$_{eff}$ for each star was determined
photometrically using the B-V values from T03 and a reddening of E(B-V)=0.22 (AT10, with an error
of $\pm$0.04).  The color-temperature relationship used is 
\small
\begin{equation}
T_{eff} = 8575 - 5222.27 (B-V)_0+ 1380.92 (B-V)_0\,^2 \\ 
        + 701.7 (B-V)_0([Fe/H]_* - [Fe/H]_{Hya}) 
\end{equation}
\normalsize
from Deliyannis et al. (2002).  We have used [Fe/H]$_{Hya}$=+0.15 (see discussion in Deliyannis et
al. 2012, in preparation) for this Hyades-based temperature scale.  From both photometric (T03) and spectroscopic
(AT10) measurements of the Fe abundance for NGC 6253, an [Fe/H] of +0.45 was used for both the
T$_{eff}$ calculation and the atmospheric metallicity.  

The surface gravities have been based on scaled-solar isochrones for [Fe/H] = +0.45, obtained from
the web interface for the Dartmouth Stellar Evolution database\\
(stellar.dartmouth.edu/models/index.html).  An appropriate age is necessary for this, and we have 
used an age of 3.0 Gyr determined from Yale-Yonsei (Yi et al. 2001; hereafter Y$^2$) isochrone
fits to the T03 photometry (AT10).  Fits with other published isochrones and age estimates by
other groups are discussed in AT10 and T03.  All of the isochrone fits give ages from 2.5
to 3.7 Gyr, which surround our choice of 3.0 Gyr.  It should be noted that the derived A(Li)
are only weakly affected by errors in log g and that the use of Dartmouth isochrones
instead of Y$^2$ isochrones for deriving log g was shown in AT10 to result in no meaningful difference.
The microturbulence values have been based on the Edvardsson et al. (1993) relation, but with some
modifications discussed in AT10.

\section{Abundance Analysis}

\subsection{Iron Abundances}

We have used 9 isolated Fe I lines in the observed Li-wavelength region (near 6708 \AA); these lines
were measured for all radial-velocity members that showed no evidence of binarity. These
measurements have been discussed in detail in AT10. Using a linearly-weighted average (in linear
space, rather than logarithmic space), they find NGC 6253 to have an extremely high metallicity at
[Fe/H]=+0.43$\pm$0.01 from their dwarf sample of 39 stars and +0.46$\pm$0.02/0.03 from their giant
sample of 18 stars. These results can be compared to the consistent analysis by Maderak et al. (2012;
hereafter M12) of 9 isolated Fe
I lines in the oxygen-triplet region. Using an identical sample of 39 NGC 6253 dwarf members and
model atmospheres, M12 found [Fe/H]=+0.445$\pm$0.014.  This shows a remarkable consistency between
the two independent measurements of different sets of Fe I lines.

Comparisons to other spectroscopic [Fe/H] measurements of NGC 6253 also show excellent consistency.
Carretta et al. (2007) derive [Fe/H]=+0.46$\pm$0.03 from spectral synthesis of a sample of 4 
red-clump stars.  Sestito (2007) derive [Fe/H]=+0.39$\pm$0.08 from equivalent-width analysis of
5 turnoff and RGB stars. The average from the equivalent-width analysis of the 4 red-clump stars in
Mikolaitis et al. (2012) is an impressively self-consistent [Fe/H]=+0.45$\pm$0.02 (rms). Lastly, 
Montalto et al. (2012; hereafter Mont12) use equivalent-width analysis to derive [Fe/H]=+0.19$\pm$0.13 from 1
main-sequence star and [Fe/H]=+0.26$\pm$0.11 from 2 red-clump stars. Although the Mont12
results are (technically) marginally consistent with all of the above studies (i.e., their results
are within roughly 2$\sigma$ of the results of all the other studies), their apparent (but
perhaps only at most marginal) deviation from all other studies merits further discussion.  First,
while their one dwarf and two giants are consistent with each other, the much larger samples of AT10
(39 dwarfs and 18 giants) are also consistent with each other, and consistent with the large sample
of 39 dwarfs of M12 that used a different spectral region.  Second, and importantly, when Mont12
impose the parameters of AT10 on their two giants, they find [Fe/H] = +0.40 and +0.43 for those two
stars, in superb agreement with AT10.  Indeed, we refer the reader to AT10 and Mont12, who discuss
various systematics in some detail, and find that most of the 
discrepancies found to date can be attributed to the choice of adopted parameters.  AT10 also note
that if differences arise among studies due to differences in the adopted parameters, they are 
usually worse among the giants than the dwarfs, especially due to the sensitivity of the giants to
assumptions for microturbulence.  Regardless, the consistency between the giants in all of these
studies is remarkable when the same set of parameters is adopted.  Furthermore, noting that the
dwarf parameters in AT10 and M12 are chosen mostly independently of the parameters used for the AT10
giants, the consistency between the dwarf studies of AT10 and M12 and between these dwarfs and
all of the giant studies (with Mont12 using the AT10 parameters) is also remarkable, and adds
support for the higher value (near +0.44).  These results show that NGC 6253 is one of the 
most-metal-rich clusters known, perhaps even the most-metal-rich cluster. 

\subsection{Lithium Synthesis}

Spectral synthesis was used to fit the Li feature at 6708 \AA\,
and the nearby blended Fe I line.  In contrast to the relatively isolated Fe I lines used
to determine [Fe/H], synthesis was seen as advantageous for measuring the strength of Li in the
super-metal-rich members of NGC 6253.  The increased strength of the neighboring metal
lines, including the partially blended Fe I line, would have made judging the continuum level and
properly fitting an equivalent width for the moderately weak Li feature challenging.  The
MOOG spectral-analysis program (Sneden 1973) was used to create and compare the synthetic and
observed spectra.  The line list for the Li region has been taken from Hiltgen (1996; see King et
al. 1997 for a discussion), which includes both fine and hyperfine structure for
the Li feature.  Table 1 includes the results of the Li synthesis for all of the member dwarfs.

The random errors in A(Li) due to spectral noise have been calculated for each star based on the 
Cayrel de Strobel \& Spite (1988) 1$\sigma$
equivalent-width error relation, as recast in Deliyannis et al. (1993), which is dependent on the
plate-scale of the telescope, the S/N of the spectrum, and the FWHM of the spectral lines.  This
provides the error of the abundance measurement based on the effects of the noise in the spectra.
This equivalent-width error and the synthetic Li measurements can be applied to our manually-created
curves of growth for the Li feature (Steinhauer et al. 2012, in preparation). This
application provides the corresponding $\sigma$(A(Li)), which is included in Table 1.  Additional
random error may be due to the judgement of continuum level, which is typically comparable to the
error due to spectral noise, but with the assistance of spectral synthesis it is much less
significant and will not be considered in this paper. 

In addition to the errors caused by observational noise, systematic errors in A(Li) need to be considered.
These are primarily due to the errors in our applied reddening because the resulting abundance
calculations are heavily dependent on T$_{eff}$.  As discussed in an section 4, we have used
E(B-V)=0.22$\pm$0.04.  With our curves of growth, we can recalculate Li abundances at the extremes
of this reddening range, and these provide our systematic errors.  Because of the small range of 
T$_{eff}$ for our stars, the systematic A(Li) error was consistently found to be  $\sim$$\pm$0.15 dex.
Although this error is larger than the $\sigma$(A(Li)), it only affects conclusions
regarding the absolute A(Li) in NGC 6253 and not the A(Li) trends seen in NGC 6253.  Additionally, 
it should be noted that this error in reddening also introduces a similar systematic error in the
[Fe/H] of the cluster.  However, a higher (or lower) [Fe/H] also would suggest a corresponding higher
(or lower) cluster initial Li due to Galactic Li production.  Hence, this systematic error on Li 
abundance may have little affect on the total Li depletion of this cluster.

Many of the observed members have undergone far too much Li depletion to make any reasonable detection.
Additionally, we prefer to consider a line as ``detected" only when its measured value is at least
3$\sigma$, where $\sigma$ is again computed according to Cayrel de Strobel \& Spite (1988).  The 
stars that did not meet this criterion are given an upper-limit Li abundance based on one of two
methods.  Method 1: The stars with lower S/N are assigned an upper limit equal to their 3$\sigma$ value. 
Method 2: The stars with higher S/N had 3$\sigma$ upper limits that were too small compared to the
strength of the nearby metal lines; therefore, they were given upper-limit equivalent widths that
were comparable to the nearby line strengths.  For both methods, the upper-limit abundances are
determined by matching the
upper-limit equivalent widths to synthetic spectra.  In Table 1, each member that did not have a
reliable $\geq$3$\sigma$ Li detection has its upper-limit abundance listed, and the calculation 
method used from the two discussed is marked.

\section{Lithium Results and Comparisons to Other Clusters}

Figure 2 is a color-magnitude diagram of all the NGC 6253 members observed in this study, and it 
shows that they are located along the turnoff.  Figure 2 also includes the non-members and binaries
with each type labeled separately.  We define two distinct regions that are marked in Figure 2: 
Region 1 is the horizontal group of more-evolved members that have little variation in V magnitude
but large variation in color.  Region 2 is the vertical group of members that have little variation
in color but large variation in V magnitude.  

For the evolved stars in region 1, Figure 3a shows the standard method of plotting A(Li) versus
T$_{eff}$, which is informative for differentiating between members of different mass.  Although
none of these members have Li detections, their 3$\sigma$ upper limits are of great
interest because they show that these higher-mass subgiants (formerly F dwarfs), which have recently
evolved off of the main sequence, have depleted their Li significantly.  Heavy Li depletion in F
dwarfs (the Li gap) is observed in all older open clusters ($\geq$650 Myr), e.g., the Hyades (Boesgaard 
\& Tripicco 1986); NGC 752 (Hobbs \& Pilachowski 1986); and Praesepe (Soderblom et al. 1993a).  
In contrast, the standard models of Li depletion (e.g., P97) predict that stars in this mass range
should have little
to no Li depletion, both in Hyades-aged clusters and in near solar-metallicity clusters as old as
NGC 6253.  Unfortunately, standard Li-depletion models do not exist for the extremely-high
metallicity of NGC 6253, but the increase in metallicity would not affect the standard Li-depletion
mechanism significantly enough to increase the standard depletion by the 2 to 3 orders of magnitude
necessary to match our results for the evolving F dwarfs in this cluster.  Additionally, the
standard model's
estimate of the additional Li depletion that occurs in evolving stars of this mass cannot explain
this magnitude of depletion (Sills \& Deliyannis 2000).  

In region 2, which has a very small T$_{eff}$ range, plotting A(Li) versus V magnitude
properly differentiates the stars (Figure 3b).  The resulting trends in A(Li) with magnitude show
a remarkably well-defined Li morphology (the Li trend with respect to T$_{eff}$, mass, or magnitude),
which is also quite similar to what is seen in Hyades-aged clusters.  The brighter
stars (14.8$\le$V$\le$15.4) show a Li gap with a broad range of Li abundances,
where most of the stars have depleted too much Li for the Li feature to be detected.  The
slightly fainter stars (15.4$\le$V$\le$16) have Li abundances that quickly increase in magnitude
and decrease in scatter.  Lastly, the Li abundances reach a peak (V$\sim$15.75), in what is
typically referred to as the Li plateau, and then quickly
drop with increasing magnitude (V$\ge$15.75) until Li is no longer detectable in the faintest
observed stars.  

Interesting conclusions can be drawn from direct comparison of the NGC 6253 Li morphology to the Li
morphologies of other clusters, but comparison to a sample of similar clusters (in either age or
metallicity) is challenging because of NGC 6253's older age and extremely high metallicity.
Nonetheless, comparisons to the Hyades (our standard reference cluster) and to M67 are fruitful.  
Whereas the Hyades is both younger and more-metal-poor ([Fe/H]=+0.135$\pm$0.005; Deliyannis et al.
2012) than NGC 6253, the solar-metallicity M67 has a more similar
age to NGC 6253 and can be used to probe effects of metallicity on Li depletion.  To improve
our analysis of age effects, we have also compared NGC 6253 to the intermediate-aged open clusters 
NGC 3680, NGC 752, and IC 4651 from AT09.  For both NGC 6253 and M67, the standard cluster
comparisons in T$_{eff}$ space are not appropriate because their member samples are composed of
evolving turnoff stars; therefore, the primary comparisons of Li abundances have been plotted
against stellar mass.  Additional comparisons are performed by plotting A(Li) against the
T$_{eff}$ that these stars would have had, or do have, at the age of the Hyades.  Both of these
transformations are based on the Y$^2$ isochrones, but for consistency our color-T$_{eff}$
relationship (discussed above) is used instead of that applied in the isochrones.

\subsection{Comparison to the Hyades}

For comparison to the younger (650 Myr) and less-metal-rich ([Fe/H]=+0.135) stars of the Hyades, 
we have used Li synthesis of our Hyades spectra from WIYN/HYDRA (Cummings et al. 2012, in preparation), our own Li
synthesis of the Hyades
spectra from Thorburn et al. (1993; provided through private communication), and the A(Li)
determined by applying the equivalent widths of Boesgaard et al. (1988), Boesgaard \& Tripicco
(1986), and Soderblom et al. (1990) to our curves of growth.  When overlap in the samples occurs, 
the final abundance is based on the highest-priority measurement, which is indicated by the order
that they have been listed above.
Reassuringly, comparison of these multiple abundance measurements shows that, except for Soderblom
et al. (1990), all of the measurement methods give consistent abundances and show no evidence for
systematic differences.  The equivalent width for the one star that we have used from Soderblom et al.
(1990) has been adjusted to a consistent scale based on the measured systematic difference between
the Boesgaard et al. (1988) and Soderblom et al. (1990) equivalent widths.  The full details
of the Hyades Li analysis and results are discussed in Cummings et al. 2012.  

Figure 4 compares the Hyades to NGC 6253 in a Li-versus-mass diagram (panel a) and a Li-versus-T$_{eff}$
diagram (panel b).  For the latter, the Y$^2$ isochrones have been used to transform the masses of the
stars to the T$_{eff}$ that they would have had at the age of the Hyades (650 Myr).  The reason for
the relative shift of features (e.g., the Li plateaus match better in Li-T$_{eff}$ space rather than
in Li-mass space) is the metallicity dependence of the mass-T$_{eff}$ relation.  Overall, 
Figure 4 shows both similarities and differences between the two clusters, and we focus first on the
similarities.  As mentioned earlier, these clusters both share the same general Li-mass morphology,
but it is important to note that the trends are shifted to lower-mass stars in the Hyades,
with heavily-depleted high-mass stars (M $>$ 1.3 M$_{\odot}$; Figure 4a), followed by a Li plateau
at slightly lower mass (M=1.3-1.2 M$_{\odot}$), then followed by a quick drop in Li abundances in
the lowest-mass stars (M $<$ 1.2 M$_{\odot}$).  Another similarity is that both clusters show
little scatter in their late-F to early-G dwarfs (6250 to 6000 K, the Li plateau).  While the
scatter appears to be larger in NGC 6253, it is primarily the result of a single Li-rich outlier
(star 709).  Therefore, there is not enough data for a reliable representation of the scatter in the
Li plateau of NGC 6253.  The relatively small scatter does suggest that Li depletion in the plateau 
of the younger Hyades, and potentially in the older NGC 6253, is primarily only affected by age, 
mass, and metallicity.  This small scatter in Li
abundances continues in lower-mass Hyades dwarfs, but for comparison we have only three detections
in dwarfs of NGC 6253 cooler than 6000 K, but those three do fall nicely on a nearly vertical Li
depletion trend with little scatter.  Another interesting similarity is that the Li plateau occurs
at the same T$_{eff}$ (6050 to 6250 K) for both clusters, though not at the same mass (see above).
This interesting feature seems to propagate all of the way to the opposite metallicity extreme, as
metal-poor (halo) dwarfs also exhibit a Li plateau in this T$_{eff}$ range (Deliyannis et al. 1990;
Ryan et al. 1996), even
as their masses continue to decrease at a given T$_{eff}$.  Lastly, the Li depletions of the cool
side of the Li gap appear to be quite similar in Li-T$_{eff}$ space, though less so in Li-mass space. 

Next, we focus on the overall differences.  The most obvious difference is that in comparison to the
Hyades, NGC 6253 has less Li across almost the entire range of observed stellar masses, with only
the abundances seen in the cool side of the Li gap being similar in both clusters.  There are
several factors that need to be considered in this comparison.  For stars in the plateau (6050 to
6250 K) and cooler, this difference in A(Li) is $\geq$0.5 dex, so the moderate systematic errors
discussed for NGC 6253 ($\sim\pm$0.15 dex, section 5.2) cannot explain this difference.  Based on
the P97 standard depletion models alone, in these stars the increase in cluster age (650 Myr to 3
Gyr) cannot explain the observed increase in Li depletion either, since the predicted increases in
Li depletion between these ages are $<$0.1 dex at either solar or Hyades metallicity.  The significant
difference in cluster metallicity, however, has the potential to be a factor: for example, the P97
models at 6000 K deplete $\sim$0.3 dex more Li at [Fe/H] = +0.15 than in comparison to [Fe/H] = -0.2.
For further comparison, P97 does not provide depletion models more metal rich than
[Fe/H] = +0.15 (corresponding to Hyades), and extrapolations to [Fe/H] = +0.45 (NGC 6253) are
ill-advised, but the comparison at lower [Fe/H] is indicative of the strong dependence that
standard Li depletion has on metallicity.

There is another potential and important complication, namely, the Galactic production of Li.  
Studying the effects of metallicity on stellar Li depletion is complicated by the
possibility that stars of higher metallicity began their lives with a higher initial-Li abundance.
Given the great difference in [Fe/H] between the Hyades and NGC 6253, the initial Li of NGC 6253
could also be significantly higher than that of the Hyades.  This would imply that the additional
Li depletion that has occurred in NGC 6253 is even greater than 0.5 dex, perhaps significantly greater
($\sim$0.8 dex?).  While it is conceivable that standard models alone could account for this (and we call
for the computation of the requisite super-high metallicity models to test these ideas!), we suspect
that non-standard mechanisms are also necessary, which may cause significant additional Li depletion
to occur in these stars between 650 Myr and 3 Gyr.

Even without precise knowledge of the cluster initial-Li abundances, we can nonetheless gain 
potentially valuable information about Li depletion from direct cluster comparisons.  We can
compare the slopes of the G-dwarf Li-depletion trends ($\sim$6100 K and below) because the
slope in each cluster is independent of the cluster's initial-Li abundance.  For early G dwarfs
(6100 to 5900 K, where we have Li detections for NGC 6253 dwarfs), the standard models of P97 predict
that all of their Li depletion occurs during the PMS only, but that the G-dwarf depletion slope
slowly increases after 650 Myr because the models evolve to slightly higher T$_{eff}$.  However, in
our comparisons the effects of T$_{eff}$ evolution can be ignored because we are using mass and 
Hyades-aged T$_{eff}$.  Additionally, the P97 models predict that an increase in metallicity causes
an increase in G-dwarf Li-depletion slope.  The much steeper depletion slope in NGC 6253 (Figure 4b)
is qualitatively consistent with this.  We cannot do a more rigorous comparison because the P97 models
go only from [Fe/H] = -0.2 up to +0.15, but the same pattern could continue to hold true at higher
[Fe/H].  However, observational evidence suggests that most (in log space) of the Li depletion in G dwarfs
(and probably the Sun) occurred after the PMS (e.g., after the age of the Pleiades, $\sim$100 Myr)
and that this main-sequence Li depletion requires non-standard mechanisms (Pinsonneault et al. 1989;
Pinsonneault et al. 1990; P97; Jeffries 2000; Deliyannis 2000; Sestito \& Randich 2005).  Since we
do not know how much standard models might increase the Li-depletion slope from Hyades's metallicity
to NGC 6253's metallicity, we must allow for the possibility that the non-standard main sequence Li
depletion could also depend on metallicity.

We can also compare the masses and Hyades-aged T$_{eff}$ of the Li-gap stars.  Beginning our
comparison in terms of mass (Figure 4a), at the low-mass side of the Li gap the significant 
depletion begins at a higher
mass in NGC 6253 (1.34$\pm$0.02 M$_{\odot}$) as compared to the Hyades (1.27$\pm$0.01 M$_{\odot}$).  
The high-mass side of the Li gap also shows different behavior in the two clusters.  The Hyades Li
abundances rise quickly at high mass (1.5 to 1.6 M$_{\odot}$) to A(Li)$\sim$3.3 dex, but at comparable
masses in NGC 6253 the Li abundances remain heavily depleted with Li \textit{upper limits} ranging
from 0.85 to 1.85 dex.  This difference between the Hyades and NGC 6253 may partly be explained by
the evolution of the NGC 6253
stars off of the main sequence, which causes additional Li depletion (via dilution) when their SCZs
increase in depth.  However, the standard model of Li depletion, as applied to evolving subgiants
in M67 (Sills \& Deliyannis 2000), predicts that our most evolved subgiants in NGC 6253 will
have only undergone roughly 1 dex of additional Li depletion beyond their main sequence abundance.
This is not enough to explain the $\geq$2 dex difference observed in the high-mass (1.5 to 1.6
M$_{\odot}$) stars of the Hyades and NGC 6253.  Since our highest mass stars in NGC 6253 are all
upper limits, we cannot state that there is no increase in A(Li) with increasing mass in this mass
range, but we can
state that there is no large and rapid increase in A(Li) in our observed high-mass NGC 6253 stars.
This may suggest that if the hot side of the Li gap ever existed, that similar to the cool side,
it occurred at a higher mass in the more-metal-rich NGC 6253.  

Detailed comparison of the Li gaps in terms of Hyades-aged T$_{eff}$ (Figure 4b) shows that the
position of the Li gap in NGC 6253 is in very good agreement with the Hyades gap.  The Li
depletion at the cool side of the Li gap begins at a similar T$_{eff}$ for both clusters ($\sim$6250
K).  Additionally, our hottest observed NGC 6253 stars and the Hyades dwarfs of similar T$_{eff}$
both now show heavily depleted A(Li), and the discussed rise in A(Li) for the hottest Hyades dwarfs
occurs at a higher Hyades-aged T$_{eff}$ than we were able to observe in NGC 6253.  Because of the
age of NGC 6253, the even higher-mass stars than those discussed in this paper are red giants.  An
additional configuration of candidate red-giant and blue-straggler members for NGC 6253 was
similarly analyzed and discussed in AT10, but all of the observed red-giant members are heavily
depleted and provided no Li detections.  

\subsection{Comparison to M67}

For comparison to the slightly older (4.5$\pm$0.5Gyr) but significantly less-metal-rich 
([Fe/H]$\sim$0.00) stars of M67, we have used our own synthesis of the spectra from Deliyannis et
al. (1997; provided through private communication with J.R. King), the equivalent width measurements
from Pasquini et al. (2008; hereafter P08), and the equivalent widths from
Jones et al. (1999).  When overlap in the samples occurs, the final abundance is based on the
highest-priority measurement, which is indicated by the order that they have been listed above.  For
consistency with our analysis of NGC 6253, we have determined our own stellar parameters by
adopting the reddening (E(B-V)=0.041$\pm$0.004) and metallicity ([Fe/H]=-0.009$\pm$0.009) from
Taylor (2007), and we have applied these equivalent widths to our Li curves of growth.  A systematic
offset of -0.3 dex was found between the resulting abundances from P08 and Jones et al. (1999), and
for consistency the Jones et al. (1999) abundances have been adjusted.  Membership in P08 has been
determined from both radial velocities
and Yadav et al. (2008) proper motions, and binarity has been tested for using radial velocities at
several different epochs.  
However, in comparison to the M67 main sequence, the photometry of 6 of the P08 likely single-star
members have very red colors for their V magnitudes, which brings uncertainty to their determined 
stellar parameters and single-star status.  These stars were also noted as possible long-period
binaries in P08; therefore, they have been removed from our final M67 Li results.  

Figure 5 compares M67 to NGC 6253 in a Li-versus-mass diagram (panel a) and a Li-versus-T$_{eff}$
diagram (panel b).  There are both interesting similarities and interesting differences.  Again, the
general trends of Li morphologies are similar: a Li gap
with a broad range of abundances, followed by a Li plateau, and a subsequent depletion trend seen in
the late F and G dwarfs.  Additionally, while the observed scatter in Li abundances is more significant
in M67, which makes defining the Li-depletion trends more challenging than in NGC 6253 or the Hyades,
we again see that in T$_{eff}$ space the two cluster gaps begin at similar T$_{eff}$ ($\sim$6250 K).  

Focusing on the Li-plateau stars and the cool side of the Li gap, we again see that they occur at a
higher mass in the more-metal-rich NGC 6253.  In comparison to NGC 6253, the cool side of the 
Hyades's Li gap occurs roughly 0.07 M$_{\odot}$ lower, and in M67 it occurs roughly 0.15 M$_{\odot}$
lower at 1.19$\pm$0.03 M$_{\odot}$, but this estimate has greater uncertainty because of M67's larger
scatter.  This pattern suggests that the Li gap occurs in higher-mass stars in higher-metallicity
open clusters, which has also been observed in the cluster comparisons of Balachandran (1995) and
of AT09.  In contrast to the Hyades, the abundances of the Li-plateau stars are comparable in both 
clusters with NGC 6253 having a A(Li) of roughly 2.55 dex and M67 showing a moderately higher
A(Li) of roughly 2.65 dex.  Considering our derived systematic errors for NGC 6253 and the Li scatter
in these stars for both clusters, this difference is not significant.  At first, this agreement may
suggest that [Fe/H] has no effect on Li depletion for stars in this mass range, but this would be
inconsistent with depletion models and the expectation from Galactic Li production models that clusters
of different [Fe/H] will have a different initial Li.  Alternatively, and perhaps more likely, this
agreement may be explained by the higher levels
of Li depletion in the metal-rich cluster being balanced out by a higher initial Li abundance for the
metal-rich NGC 6253.  We can reach this conclusion even without detailed
knowledge of the additional non-standard depletion mechanisms because for this conclusion
to be incorrect, it would require the additional-depletion
mechanisms to become significantly weaker as the metallicity increased, but there have been no
observations to suggest this and it is not seen in any of the current models of 
non-standard Li depletion.  Any non-standard depletion occurring between the ages of 3 and 4.5
Gyr may also play an important role in this comparison and our conclusion, but at the temperature
range near the Li plateau, the difference in additional depletion is not likely significant enough to
greatly change our conclusion that the metal-rich NGC 6253 had a larger initial Li than the 
solar-metallicity M67.  This higher initial Li is also consistent with observations of field dwarfs
and models of Galactic Li production (e.g., Ryan et al. 2001; Travaglio et al. 2001).  However, it
is of interest to note that the Galactic Li production models of Fields \& Olive (1999a, 1999b) 
suggest that at super-solar metallicities the effects of stellar Li depletion begin to have an 
impact on the Li abundance in the ISM and may flatten or even reverse the correlation of
initial Li and Fe.  Such a reversal would suggest that NGC 6253 may have had a comparable or even
smaller initial Li than either M67 or the Hyades.  Nonetheless, we remind the reader that no 
Galactic Li production models have yet been able to even produce enough Li to account for
the meteoritic Li abundance of 3.31, and so there exists (at least) a corresponding uncertainty as 
to how Galactic Li procedes at super-solar metallicities.

There are several remaining differences between NGC 6253 and M67 that are important.  First,
the cooler ($\leq$6050 K) stars of NGC 6253 are much more heavily depleted than the stars of 
similar T$_{eff}$
in M67.  This was also seen in the Hyades and is, again, qualitatively consistent with P97 models.
This difference in Li abundances is even more dramatic if, as previously discussed, NGC 6253 had
a higher initial Li.  Additionally, independent of initial Li
and as predicted by the models, the more-metal-rich NGC 6253 has a steeper G-dwarf depletion trend.
Second, there is a much larger scatter in A(Li) among the lower-mass stars
of M67 than in NGC 6253.  Further evidence for the scatter in the M67 Li abundances is shown by the
comparison to the Sun (Figure 5a), which has an age and metallicity consistent with M67.  
However, the scatter, which appears to become more significant with decreasing mass, is
potentially not seen in NGC 6253 due to the limited number of Li detections in its lower-mass members.

\subsection{Comparisons to Additional Clusters}

Here, we make additional comparisons to the three intermediate-aged clusters of slightly varying 
metallicity: NGC 3680, NGC 752, and IC 4651, using consistently determined Li abundances
from published equivalent widths.  (In AT09, we presented results for NGC 3680, and compared this
cluster to NGC 752 and IC 4651.)
The equivalent widths used for NGC 752 were taken from Hobbs \& Pilachowski (1986), Pilachowski \&
Hobbs (1988), and Sestito et al. (2004).  The equivalent widths used for IC 4651 were taken
from Pasquini et al. (2004), Balachandran et al. (1991), and the unpublished data from S.
Balachandran (1990, private communication).  When stars were taken from more than one source, 
comparisons showed that in all but one case there were no significant differences in equivalent
widths.  The higher S/N measurement was used in the single discrepant case, and the average of the
equivalent widths was used in all other cases.  The properties of the three intermediate-aged
clusters and all others discussed in this paper are summarized in Table 2 (see the detailed
discussions of reddening in AT09).

Comparisons are made to NGC 6253 in Figures 6a and 6b for NGC 3680, in Figures 7a and 7b for NGC
752, and in Figures 8a and 8b for IC 4651.  All three clusters show similar general Li morphologies.
Particularly, all show cool dwarf (T$_{eff}$ $\geq$ 6000 K) Li depletion (except IC 4561, for which
cool-dwarf Li data do not exist), a Li plateau (near 6000 to 6250 K), and a Li gap with its cool side at
similar T$_{eff}$ as NGC 6253 ($>$6250 K).  Owing to their younger ages, the three clusters also
continue to have hotter stars on the main sequence that exhibit the hot side of the Li gap (larger
Li at T$_{eff}$ near 6800 to 6900 K), Li abundances with a (sometimes large) scatter around 3.0 for main
sequence stars hotter than 7000 K, and a very steep (nearly vertical!) drop of Li abundances for
stars evolving off of the main sequence (at roughly 7700 K).  
As was true for the other clusters, these general features are shifted in Li-mass space, so that, for
example, the Li plateaus (and cool sides of Li gaps) no longer coincide.  As we discuss in section
7, these shifts seem to be related, once again, to the metallicity dependence of the T$_{eff}$-mass
relation.  As we did for NGC 6253, the Hyades, and M67, we determine the mass of the cool side of
the Li gaps to be about 1.17$\pm$0.02, 1.19$\pm$0.025, and 1.29$\pm$0.025 M$_{\odot}$ for NGC
3680, NGC 752, and IC 4651, respectively.

\subsection{Comparison Summary}

We have compared our Li abundances for NGC 6253 to those of the Hyades, NGC 752, IC 4651, NGC 3680,
and M67, which span a broad range of age and [Fe/H] (see Table 2).  All of these clusters
exhibit a Li gap, and we
have found the rather striking result that the cool side of the Li gap is observed at very similar
T$_{eff}$ in all of these clusters, irrespective of their substantial range in age (0.65 to 5 Gyr)
and in [Fe/H] (-0.1 to +0.45).  (A similar conclusion was reached by Balachandran (1995) 
using more limited data in a subset of four of these clusters.)  However, the cool side of the Li
gap does depend on mass, which seems to be a reflection of how the T$_{eff}$ of main-sequence F-dwarf
mass depends on metallicity.

This dependence has been quantified as a correlation between the Li gap's central mass, M$_{dip}$, 
and [Fe/H] by Chen et al. (2001) in field dwarfs with a broad range in [Fe/H], and by AT09 using
a subset of the six clusters (Hyades[+Praesepe], NGC 752, IC 4651, and NGC 3680; thus excluding the
solar-metallicity M67 and the super-metal-rich NGC 6253).  To within the errors, both studies agree
on the slope and intercept of the linear M$_{dip}$/[Fe/H] relation, underscoring a potentially important
similarity between field and cluster dwarfs.  We hereby extend this type of analysis to include M67
and NGC 6253; however, due to the older ages of these clusters, the stars on the hot side of the Li
gap have evolved off of the main sequence.  So, rather than define a central mass for the Li gap, 
which is poorly defined in these two clusters, we define a mass for the cool side of the Li
gap, M$_{cs}$, for all six clusters.  Figure 9 shows M$_{cs}$ versus cluster [Fe/H] for all six clusters.  
We have fit the data two different ways, with a linear relation (M$_{cs}$/M$_{\odot}$=1.21+0.35[Fe/H]; 
dot-dash line) or with a quadratic relation (M$_{cs}$/M$_{\odot}$=1.21+0.54[Fe/H]-0.55[Fe/H]$^2$; 
dashed curve).  This new linear relation has a slope with [Fe/H] of 0.35$\pm$0.05 M$_{\odot}$/dex, 
consistent with 0.4$\pm$0.2 from AT09 and comparable to 0.28 from Chen et al. (2001), but Chen et 
al. (2001) gave no uncertainty estimate.  However, with the addition of NGC 6253, we see the
possibility that the relation
turns over smoothly and flattens out as metallicity increases, as suggested by the dashed curve, but
it should be noted that the quadratic term in this dashed curve is significant at only the 2$\sigma$
level.  Additionally,
one might wonder whether the turn over of M$_{cs}$ in NGC 6253 is due to its advanced age; that is,
whether its M$_{cs}$ has evolved to lower mass as the cluster has aged, but such an evolution is 
\textit{not} apparent in the other clusters, one of whom (M67) is at least as old as NGC 6253.  
However, all of these other clusters are substantially more-metal-poor than NGC 6253, so
such an evolution cannot altogether be ruled out.  Nonetheless, it is quite possible that this 
effect is purely a result of NGC 6253's high metallicity, alone.  As a further comparison, and as we
suggested earlier, this M$_{cs}$-versus-[Fe/H] correlation may purely be a reflection of how the T$_{eff}$ of
main-sequence F-dwarf mass depends on metallicity.  Therefore, in Figure 9 we have also compared
the data to the mass and [Fe/H] relation for Hyades-aged stars at a T$_{eff}$ of 6250 K (solid
curve).  Remarkably, the data and this relation are consistent, and this may suggest that
the correlation of Li-gap mass and [Fe/H] may solely be a result of the relation between mass and
[Fe/H] in dwarfs at this temperature.

\section{Constraints on the Origin of the Lithium Gap}

The Li observations in NGC 6253 can potentially test and constrain the variety of mechanisms that
have been proposed to explain the origin of the Hyades Li gap.  These mechanisms include mass loss
(Schramm et al. 1990; simply that the star loses the Li preservation region), diffusion (Richer \&
Michaud 1993; gravitational settling and thermal diffusion forms the cool side of the gap whereas
radiative acceleration forms the hot side of the gap), and various forms of slow mixing, including
rotationally-induced mixing (Pinsonneault et al. 1990; mixing due to angular-momentum loss causes
surface Li depletion).  It is clear from our measurements that the subgiants of NGC 6253
were part of the Li gap.  The deepening SCZs of subgiants evolving out of the Li gap can reveal
the internal Li profiles, and can thus constrain or even point to the dominant Li depletion
mechanism(s) that create the Li gap (Deliyannis et al. 1996; Sills \& Deliyannis 2000).  A very
steeply declining Li-T$_{eff}$ relation would favor mass loss, a Li-T$_{eff}$ relation declining
less steeply would favor rotational mixing, and an \textit{increasing} Li-T$_{eff}$ relation would
favor diffusion.  This increasing relation would be caused by the dredge-up of Li diffused but not
destroyed during the main sequence.  As was the case with M67 subgiants (Sills \& Deliyannis 2000),
our (region 1) NGC 6263 subgiants show no evidence of dredged-up Li; therefore, they argue against
diffusion as the cause of the Li gap in NGC 6253.  Conversely, the declining subgiant Li upper
limits in NGC 6253 (Figure 3a) are consistent with both mass loss and
rotational mixing, but the lack of detections prevents us from drawing any stronger conclusions. 

That the Hyades-aged T$_{eff}$ of the cool side of the Li gap is independent of age and metallicity
(using six clusters with age and metallicity in the ranges of 0.65 to 4.5Gyr and -0.1 to +0.45 dex,
respectively) also provides constraints for the physical origin of the Li gap:

\textbf{Mass Loss:} Schramm, Steigman, \& Dearborn (1990) discuss how mass loss could create the
Hyades Li gap.  Significant mass loss of at least 7 x 10$^{-11}$ M$_{\odot}$ yr$^{-1}$, a factor
of 10,000 higher than that of the Sun, would be required to shed enough of the star's outer envelope
for the SCZ to come in contact with the Li destruction region in the interior.  At this mass-loss
rate, these gap stars should cool and extend the gap to a cooler T$_{eff}$ as they age, but this
is not seen in the much older NGC 6253.  Additionally, significant mass loss should completely
strip all of the gap stars of their Li by the age of NGC 6253, but four of the cooler gap stars
still show clear Li detections in NGC 6253.

\textbf{Diffusion:} Richer \& Michaud (1993) analyze the effects of Li diffusion and gravitational
settling.  In G dwarfs, downward thermal diffusion and gravitational settling is negligible, but
these effects increase with T$_{eff}$ as the SCZ become shallower, creating the cool side of the Li
gap, and culminating in severe Li depletion in mid-F dwarfs.  There, the SCZ is shallow enough so
that Li retains an electron, so for hotter T$_{eff}$, Li is radiatively accelerated into the even shallower
SCZ.  This effect creates the hot side of the Li gap and predicts Li overabundances near 7000K (see
below).  However, for the age of the Hyades, the models predict a Li gap with a width of only of
$\sim$200 K, which is roughly 2 to 3 times narrower than that observed.  Their models also predict
that as stars age, the Li gap evolves with time to cooler T$_{eff}$.  Additionally, as discussed
above, these effects would cause the Li abundances of the evolving stars in the gap to greatly
increase.  Neither of these predictions are observed in the Li abundances of NGC 6253.  

One additional prediction of interest from these diffusion models is that a small range of stars at
the hot side of the Li gap should be super-Li rich because Li is radiatively accelerated upward
into their extremely thin SCZs.  Rapid rotation may inhibit this diffusion process, but a relatively
slow-rotating super-Li-rich star has been observed in the Hyades-aged open cluster NGC 6633, and it
is at the hot side of the cluster's Li gap where the diffusion models predict these stars should exist. 
(Deliyannis et al. 2002).  (It should be noted that planetesimal accretion has also been
proposed as an
explanation for this super-Li-rich star (Laws \& Gonzalez 2003; Ashwell et al. 2005).)  If additional
super-Li-rich stars in this key T$_{eff}$ range can be found, this would strongly
suggest that while diffusion alone cannot produce the observed Li gap, it may still play an 
important role in the Li abundances of slower-rotating stars near the Li gap.

\textbf{Mixing:} Another mechanism that may lead to significant Li depletion in the stars of the
Li gap is rotationally-induced mixing.  One approach to modeling the effects of rotation is based on the
ideas of Endal \& Sofia (1976, 1978, 1981) as updated by Pinsonneault et al. (1988), which consider a
number of rotationally-induced instabilities.  Of relevance here, stars that have at least a minor
SCZ will begin to efficiently lose angular momentum from their surface as they age because of the
interaction of their magnetic field with their stellar wind.  This surface angular momentum
loss causes the surface layers to rotate more slowly than the interior of the star,
which can lead to secular-shear instabilities (Zahn 1987), which cause the transfer of angular momentum from
the interior to the surface and also induce mixing in the outer layers of the star.  Stars rotating
faster lose more angular momentum and they lose it more efficiently, and should accordingly induce
greater internal mixing and Li depletion.  (Meridional circulation was also considered in these models
but was argued to result in negligible mixing.)  Consistent with the predictions of rotational-induced
mixing together with the tidal-circularization theory of Zahn \& Bouchet (1989), the observations
of short-period 
tidally-locked binaries (SPTLBs) in the young ($\sim$100 Myr) Pleiades have comparable A(Li) to
the other stars of similar mass, but SPTLBs in the older Hyades and M67 have higher A(Li) than the
other stars of similar mass in their cluster (Ryan \& Deliyannis 1995).  This can be explained
by rotationally-induced mixing because the Li abundances in STPLB stars will not be as significantly
affected by rotation; STPLBs spin down very rapidly at a young age before their interiors reach a
high enough temperature to destroy Li and thus are predicted \textit{not} to suffer the 
rotationally-induced Li depletion that single stars of the same T$_{eff}$ will experience later, when 
those stars spin down.  The STPLBs have similar A(Li) in comparison to single Pleiades dwarfs at the same
T$_{eff}$ because the Pleiades are so young ($\sim$100), so they have not yet lost enough angular
momentum from their surface to induce mixing and deplete their surface Li.  In contrast,
the comparable single dwarfs of the Hyades and M67 are old enough to have lost significant surface angular
momentum and have experienced significant rotationally-induced depletion of Li.

Pinsonneault et al. (1990; hereafter P90) applied this approach of modeling rotation to A, F, G,
and K dwarfs, and their models do indeed predict a Li gap in F dwarfs, main sequence Li depletion
in G dwarfs, and a Li plateau between these two features (late-F/early-G dwarfs).  In late G dwarfs,
the SCZ is almost as deep as the Li preservation region, so rotation needs to do only a little work
(in the radiative layers in between) to deplete surface Li.  As the SCZ becomes more shallow with
higher T$_{eff}$, the radiative region increases, and the observed Li abundances rise.  However, 
stellar initial angular momenta also rise, resulting in more Li depletion, and thus the prediction 
of the Li plateau and cool side of the Li gap.  Mid-F stars coincide with the break in the ``Kraft curve" 
(stellar rotation rates versus spectral type).  Stars hotter than this no longer spin down as they
evolve, so in this context, the hot side of the Li gap is simply the absence of rotationally-induced
mixing.  Another positive feature of these models is that the Li gap does not evolve to lower T$_{eff}$,
consistent with our cluster comparisons.  The scatter in A(Li) in the gap and Li plateau is
interpreted as a reflection of the range of initial angular momenta (J$_o$) observed in young stars;
however, the relatively tight Li-T$_{eff}$ relation for G dwarfs would require a smaller-than-expected 
range in J$_o$ and the J$_o$ would have to depend on mass in just the right way.  Also, these models 
appear to deplete too much Li in the Li plateau region.

Models that take into consideration both rotationally-induced mixing and diffusion (Chaboyer et al.
1995a and 1995b) have focused on the G-dwarf Li-depletion problem, and in this regard they are more
complete than the P90 models, which only consider rotationally-induced mixing.  These models also
employ improved opacities, which could result in deeper G-dwarf SCZs.  The effects of
rotation and diffusion can partly counterbalance each other.  Higher levels of 
rotationally-induced mixing can inhibit diffusion in the fastest of rotators while an increased
gradient in mean molecular weight, primarily due to the diffusion of $^4$He, can inhibit rotational
mixing.  These models do a better job of reproducing the shape of the Li-T$_{eff}$ relation in the
Hyades G dwarfs, if J$_o$ does not depend on mass, and they also have less scatter than the P90
models, though perhaps still too much.  However, these models predict too-rapid an internal-rotation
rate for the Sun.  Chaboyer et al. (1995a) discuss how additional angular momentum transport 
mechanisms like magnetic fields or internal gravity may play an important role in better matching
both the observed Hyades G-dwarf Li scatter and the solar internal-rotation rate.

Models that take into consideration (a different but related formulation of) rotationally-induced
mixing, diffusion, and internal gravity waves produced at the base of the surface convection zone
(Charbonnel \& Talon 2005; hereafter CT05; Talon \& Charbonnel 2005; Talon \& Charbonnel 2003) can
recreate more successfully the observed A(Li) in both the gap and in G dwarfs.  These gravity waves
becomes important when the SCZ becomes more significant in dwarfs near the cool side of the Li gap
and cooler.  Gravity waves provide another efficient mechanism for transporting angular
momentum from the faster rotating interior to the exterior, and this decreases the transfer through
secular-shear instabilities and meridional circulation, another rotationally-induced mechanism
considered in these models.  Unlike the other two transfer mechanisms, however, gravity waves
do not produce significant mixing with their transport of angular momentum.  For stars cooler than
the Li gap, where gravity waves become important, this both decreases the predicted additional
depletion caused by rotation and greatly decreases the predicted variation in Li depletion
due to varying initial angular momentum.  Consistent with the cluster comparisons in this paper, these
models do not cause the position of the gap to evolve with age.  Additionally, these models
provide a mechanism that greatly slows the rotation rate of a star's interior, which is necessary
to match the flat rotation profile observed in the Sun through helioseismology (Brown et al. 1989;
Kosovichev et al. 1997; Couvidat et al. 2003).  At the same time,
enough core angular momentum must survive to explain, simultaneously, the rapid rotation observed
in horizontal branch stars, but very slow rotation in their (metal-poor) turnoff progenitors.  The
models of Pinsonneault et al. (1991) meet these (and other) rather stringent constraints; perhaps 
an inference is that gravity waves are not as effective at removing core angular momentum in main
sequence halo dwarfs near the turnoff (roughly 6300 K) that likely have shallow SCZs.

For solar-type stars, the CT05 models predict that variation among the initial angular momentum
values has only a minor effect on the Li abundances and will not produce a significant scatter by
the age of the Hyades, qualitatively consistent with the Hyades data (Figure 4).  However, these
minor effects due to variation in rotation will gradually increase the Li scatter with time, and the
scatter will become significant by the age of M67, qualitatively consistent with the large scatter 
in the M67 data (Figure 5).  For comparison of the Hyades and NGC 6253, the number of observed
plateau stars in NGC 6253 are limited, and the scatter cannot reliably be based on the single 
relatively Li-rich star 709.  However, NGC 6253's Li
scatter does not appear to be as significant as that observed in M67, but this may be because
M67 is approximately 1.5 Gyr older; the model of CT05 predicts that in solar-type stars the Li
depletion will continue during this time and that the observed Li scatter will increase between
these two ages.  This additional non-standard Li depletion in solar-type stars between these ages
may also play an important role in several of our other comparisons of NGC 6253 and M67, as has
been discussed, but based on the models for solar-type stars in CT05, this effect would not
be significant enough to change any of our conclusions.  The effects of varying 
metallicity have not been considered in the models of CT05, so we do not know whether these
mechanisms would produce Li gaps of consistent T$_{eff}$ at such different metallicities.  

\section{Summary}

NGC 6253's Li morphology is very similar to the younger and less-metal-rich Hyades and the
similarly-aged but solar-metallicity M67.  Comparisons to the intermediate-aged and 
subsolar-metallicity NGC 3680 and NGC 752 clusters and to the less-metal-rich IC 4651 also show
similar Li trends.  It is of interest that the abundances of the Li-plateau stars are comparable in NGC
6253 and M67.  This does not necessarily suggest that Li depletion is independent of metallicity in
these stars, but perhaps that at this age the more significant Li depletion
of the metal-rich NGC 6253 may have been balanced by a higher initial Li abundance.  This positive
correlation of initial Li and [Fe/H] is predicted by models of Galactic Li production.  Comparisons
of the G-dwarf Li-depletion trends in NGC 752, NGC 6253, M67 and the Hyades are also consistent with
the metallicity-dependence of the Li depletion of the P97 standard models, where the more-metal-rich
NGC 6253 has a steeper depletion trend than the other three clusters.

The mid-F Li gap requires a Li depletion mechanism (or mechanisms) beyond those included in standard
stellar evolution theory.  Analyzing Li-gap stars that are also turnoff stars can help
to test models of the additional depletion mechanisms that could create the Li-gap.
For example, models with diffusion predict that subgiants evolving out of the Li gap should experience
a sharp rise in their surface Li abundances, but an increase in A(Li) is not detected in the
evolving stars of NGC 6253; a similar result was found in M67, where the Li in evolving subgiants
declines sharply, rather than rises (Sills \& Deliyannis 2000).
Our comparison of the masses of the Li-gap stars in all six clusters suggests that in more 
metal-rich clusters the gap occurs in higher-mass stars.  This is consistent with the results from
Balachandran (1995) and AT09, but the addition of the super-metal-rich NGC 6253 suggests that this
correlation with gap position and metallicity may not be linear.  In fact, this correlation may
simply be a reflection of how the T$_{eff}$ of the Li-gap mass depends on metallicity: a comparison of
these clusters at their Hyades-aged T$_{eff}$ shows that the positions of their Li gaps occur at nearly 
identical T$_{eff}$ -- across a wide range in age and metallicity.  We underscore the finding that the 
Li gap does not appear to evolve with age.  This provides a strong test for a variety of the mechanisms
that could create the Li gap.  Models with mass loss and diffusion both predict that the Li gap will
evolve to lower-mass dwarfs as a cluster ages, so our result provides strong evidence against these
mechanisms being the primary cause of the Li gap.  Models with rotationally-induced mixing are more
successful and predict that the Li gap's position does not evolve with age, but they also predict
stronger depletion and a larger scatter than that observed in plateau stars.  Models
that combine the effects of rotationally-induced mixing and diffusion decrease this predicted
scatter, but it is still more significant than that observed.  Models that combine 
rotationally-induced mixing and transport of angular momentum through gravity waves can match the
observations more successfully.  However, we call on future models to consider the effects of 
varying the metallicity.

\section*{Acknowledgements}
We gratefully thank the National Science Foundation for supporting this project under grant 
AST-0607567, and we are also thankful for the financial support from Fondo GEMINI-CONICYT 32100008
and from the Chilean BASAL Centro de Excelencia en Astrof\'isica y Tecnolog\'ias Afines (CATA) 
grant PFB-06/2007.

\nocite{*}
\renewcommand{\bibname}{References}
\bibliographystyle{apj}
\bibliography{n6253bib}{}

\begin{figure}[htp]
\begin{center}
\includegraphics[scale=0.8]{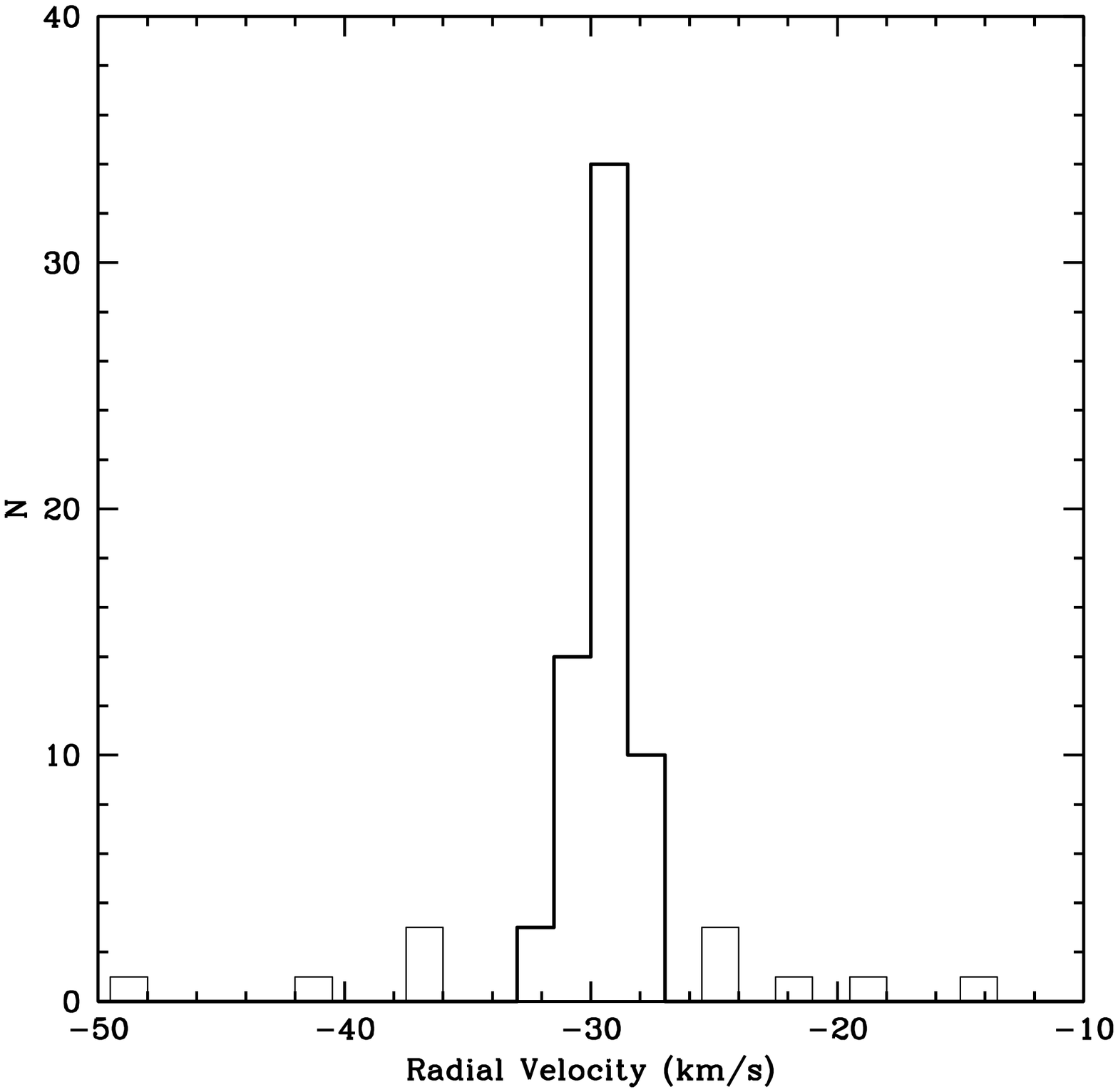}
\end{center}
\caption{The radial-velocity histogram for all observed stars.  There is a 
clear peak near -30 km/s, which represents the cluster radial velocity.  The region in bold
represents the radial-velocity range selected to contain likely cluster members.  Based on the radial
velocities from AT10.}
\end{figure}

\begin{figure}[htp]
\begin{center}
\includegraphics[scale=0.65]{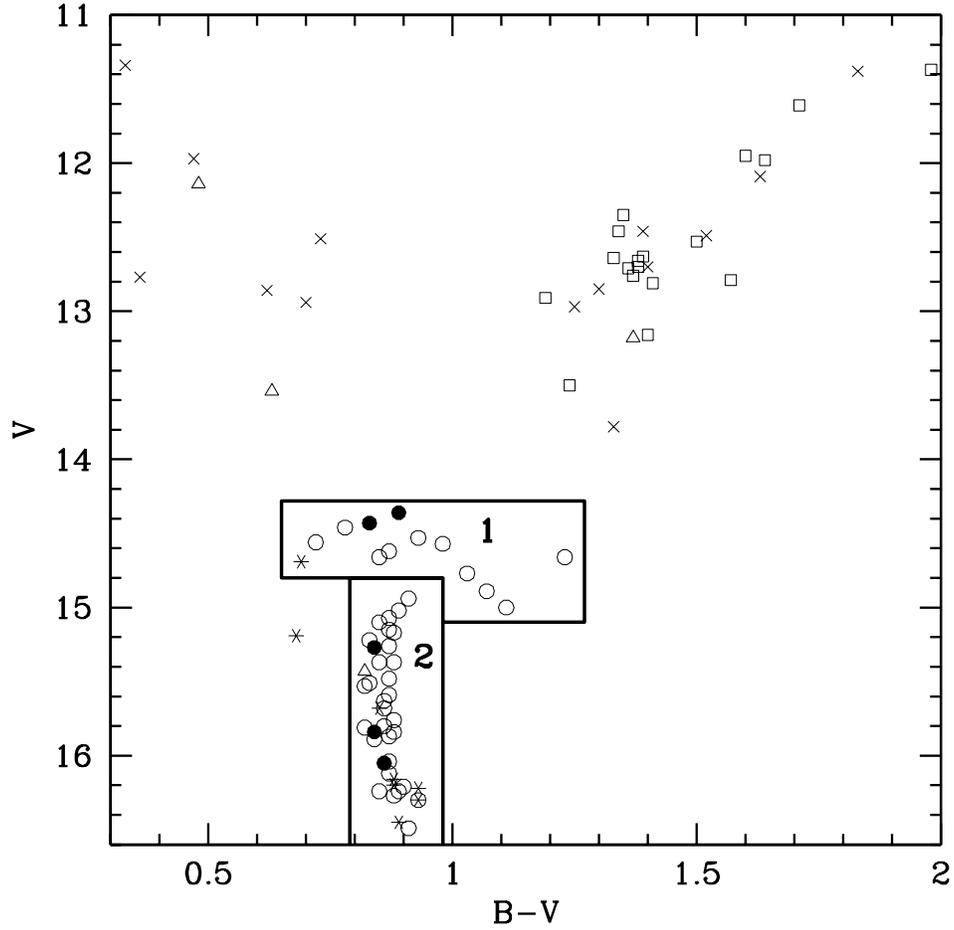}
\end{center}
\caption{The color magnitude diagram for the stars observed in our configuration, which are
the stars in region 1 and 2.  The turn
off is clearly seen with the more-evolved stars designated in region 1 and the less evolved
stars designated in region 2.  All observed members, non-members, and binaries are included
on this diagram, where open circles are turnoff members, filled circles are binary
turnoff stars, open squares are red-giant members, triangles are red giants and blue stragglers
of uncertain membership, and asterisks and crosses are both non-members.}
\end{figure}

\begin{figure}[htp]
\begin{center}
\includegraphics[scale=0.86]{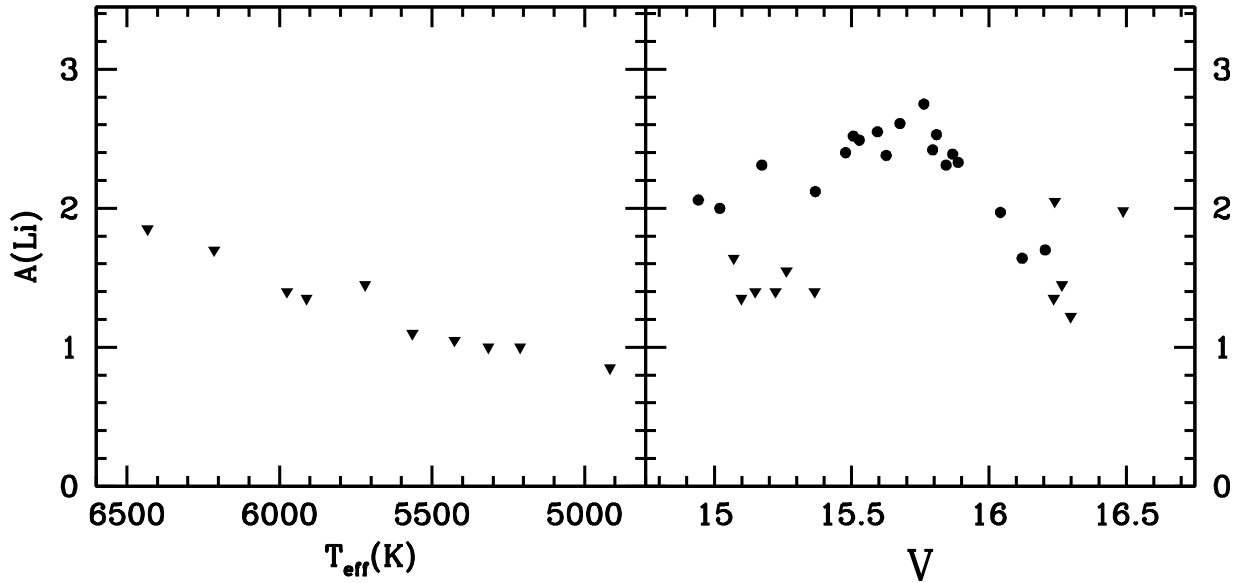}
\end{center}
\caption{a) The evolved stars in region 1 have their Li upper limits plotted versus
their T$_{eff}$.  None of the stars in this region have had a Li detection above the noise, and so
the 3$\sigma$ upper limits are given (solid inverted triangles).  They show that a significant amount
of depletion has occurred in these stars, which is quite inconsistent with standard models but 
agrees well with observations of other moderately-old to old clusters.  b) The lower-mass stars in
region 2 have their Li abundances (solid circles) and upper limits plotted versus magnitude.  The
stars clearly exhibit the general observed Li abundance morphology seen in older clusters.}
\end{figure}

\begin{figure}[htp]
\begin{center}
\includegraphics[scale=0.86]{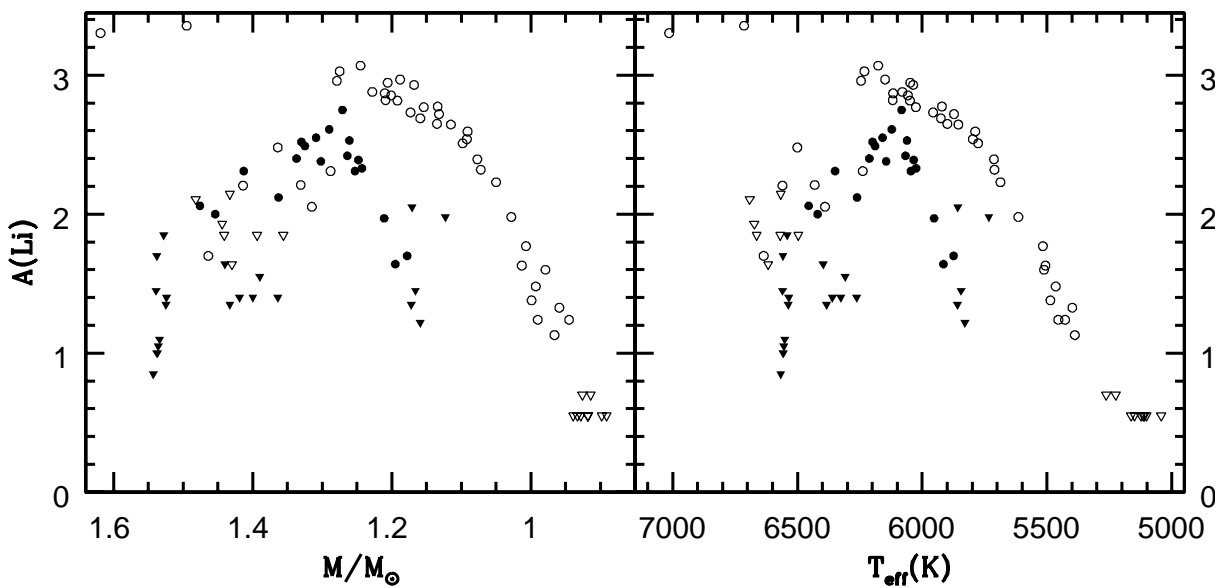}
\end{center}
\caption{a) The stars in both region 1 and 2 of NGC 6253 (solid circles and triangles) are
plotted in comparison to the Hyades (open circles and triangles) by mass.  The clusters 
show similar Li morphologies, but NGC 6253 shows less Li across the
entire mass range and also shows a more significant G-dwarf depletion trend.  Both of these factors
can only be partly explained by age, and the far greater metallicity of NGC 6253 accounts for the
more significant depletion, on top of the likely greater initial Li abundance. b) The same data are
compared by T$_{eff}$ with the NGC 6253 data transformed to a Hyades-aged T$_{eff}$ scale.  Note 
that the cool sides of their gaps occur at a similar T$_{eff}$.}
\end{figure}

\begin{figure}[htp]
\begin{center}
\includegraphics[scale=0.86]{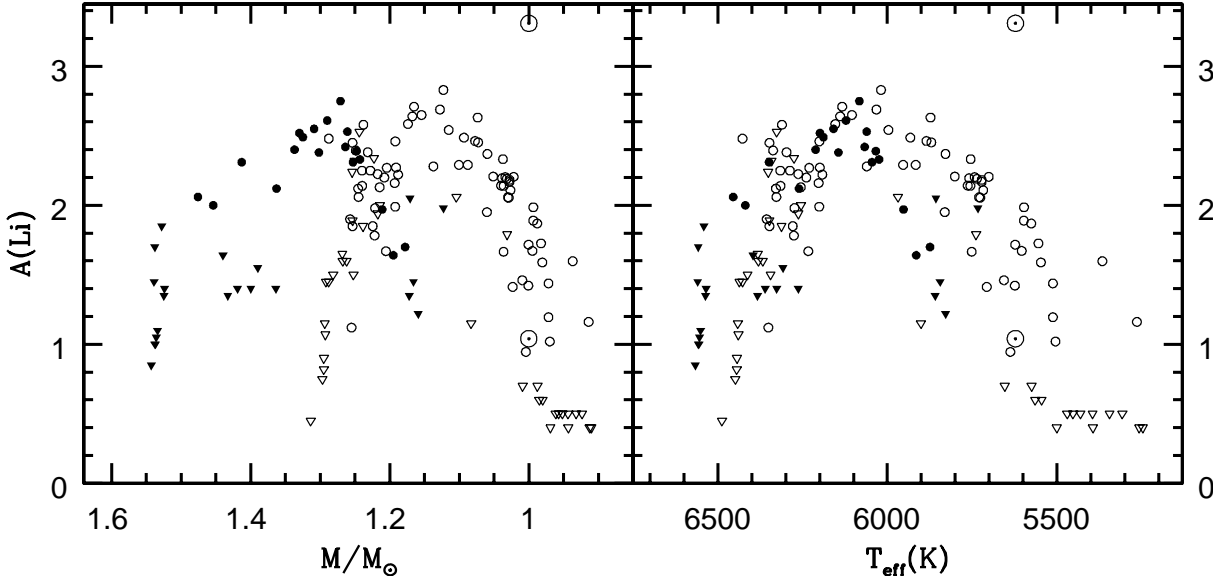}
\end{center}
\caption{a) Our entire sample of stars in NGC 6253 (solid circles and triangles) are plotted in
comparison to M67 (open circles and triangles) by mass.  The
clusters show very similar Li morphologies and even show consistent Li abundances at the Li plateau.
NGC 6253 shows a more significant G-dwarf depletion trend than M67.  The meteoritic Li and current
solar Li (upper and lower $\odot$) is shown for comparison.
b) The same data are compared by the T$_{eff}$ these stars and the Sun would have had at the age of
the Hyades.  In both figures note the greater scatter in the M67 Li abundances.}
\end{figure}

\begin{figure}[htp]
\begin{center}
\includegraphics[scale=0.86]{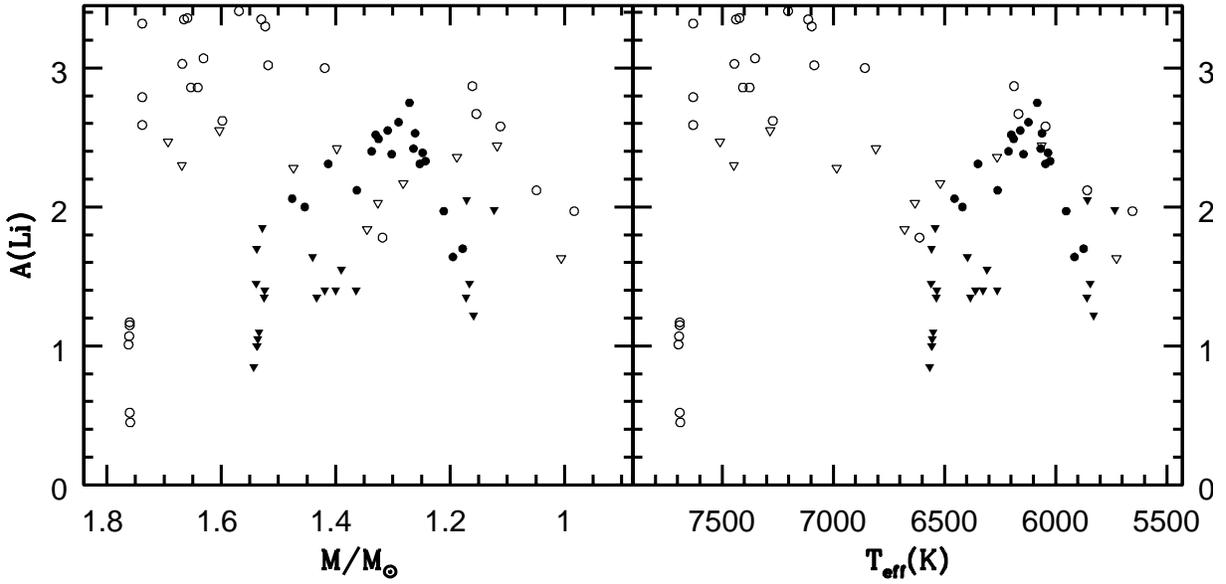}
\end{center}
\caption{a) Our entire sample of stars in NGC 6253 (solid circles and triangles) are plotted in
comparison to NGC 3680 (open circles and triangles) by mass.  Note the difference in position of
their gaps.
b) The same data are compared by the T$_{eff}$ these stars would have had at the age of the Hyades.
Note that the cool sides of their Li gaps are at a similar T$_{eff}$.}
\end{figure}

\begin{figure}[htp]
\begin{center}
\includegraphics[scale=0.86]{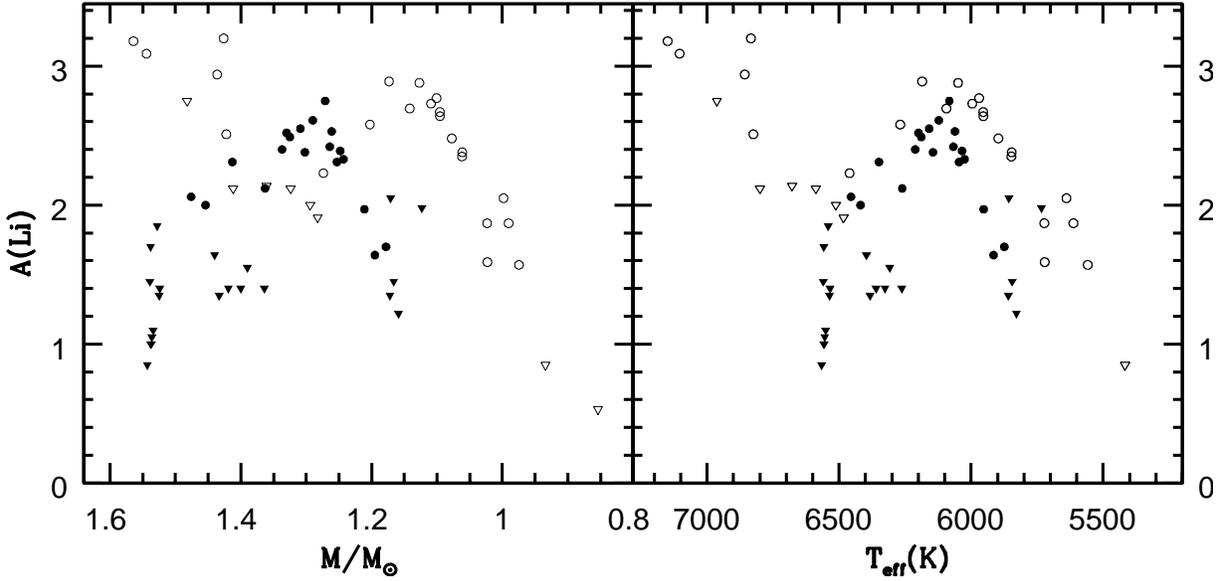}
\end{center}
\caption{a) Our entire sample of stars in NGC 6253 (solid circles and triangles) are plotted in
comparison to NGC 752 (open circles and triangles) by mass.  NGC 6253 shows a more significant 
G-dwarf depletion trend than NGC 752.  Note the difference in position of their gaps.
b) The same data are compared by the T$_{eff}$ these stars would have had at the age of the Hyades.
Note that the cool sides of their Li gaps are at a similar T$_{eff}$.}
\end{figure}

\begin{figure}[htp]
\begin{center}
\includegraphics[scale=0.86]{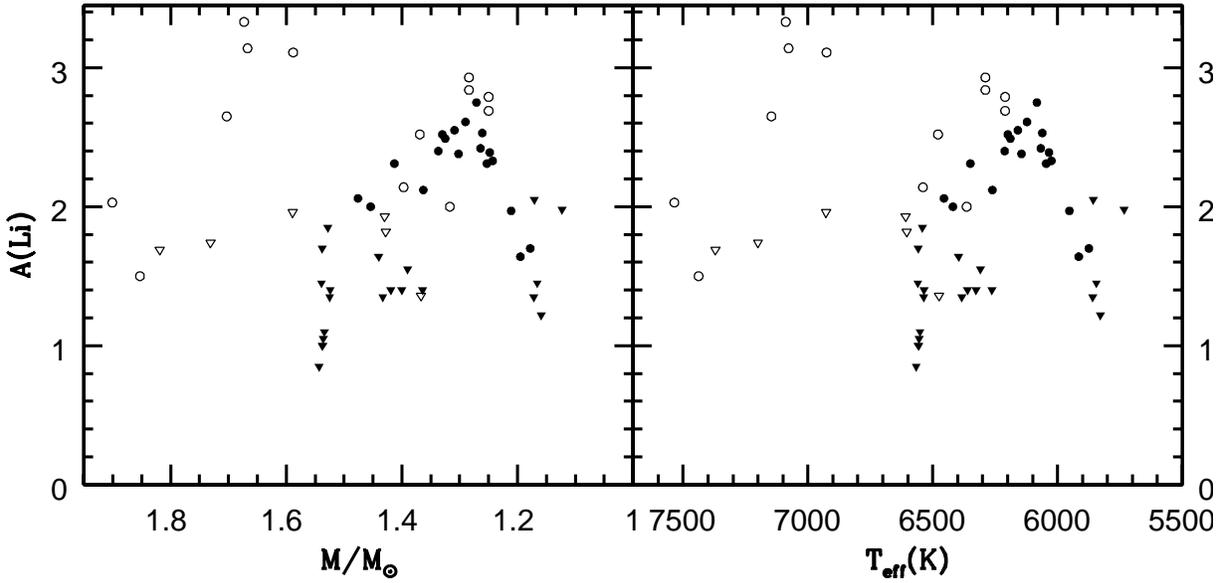}
\end{center}
\caption{a) Our entire sample of stars in NGC 6253 (solid circles and triangles) are plotted in
comparison to IC 4651 (open circles and triangles) by mass.  Note the difference in position of
their gaps.
b) The same data are compared by the T$_{eff}$ these stars would have had at the age of the Hyades.
Note that the cool sides of their Li gaps are at comparable T$_{eff}$, differing by $\sim$60 K, but
this is consistent when considering the large errors for IC 4651.}
\end{figure}

\begin{figure}[htp]
\begin{center}
\includegraphics[scale=0.8]{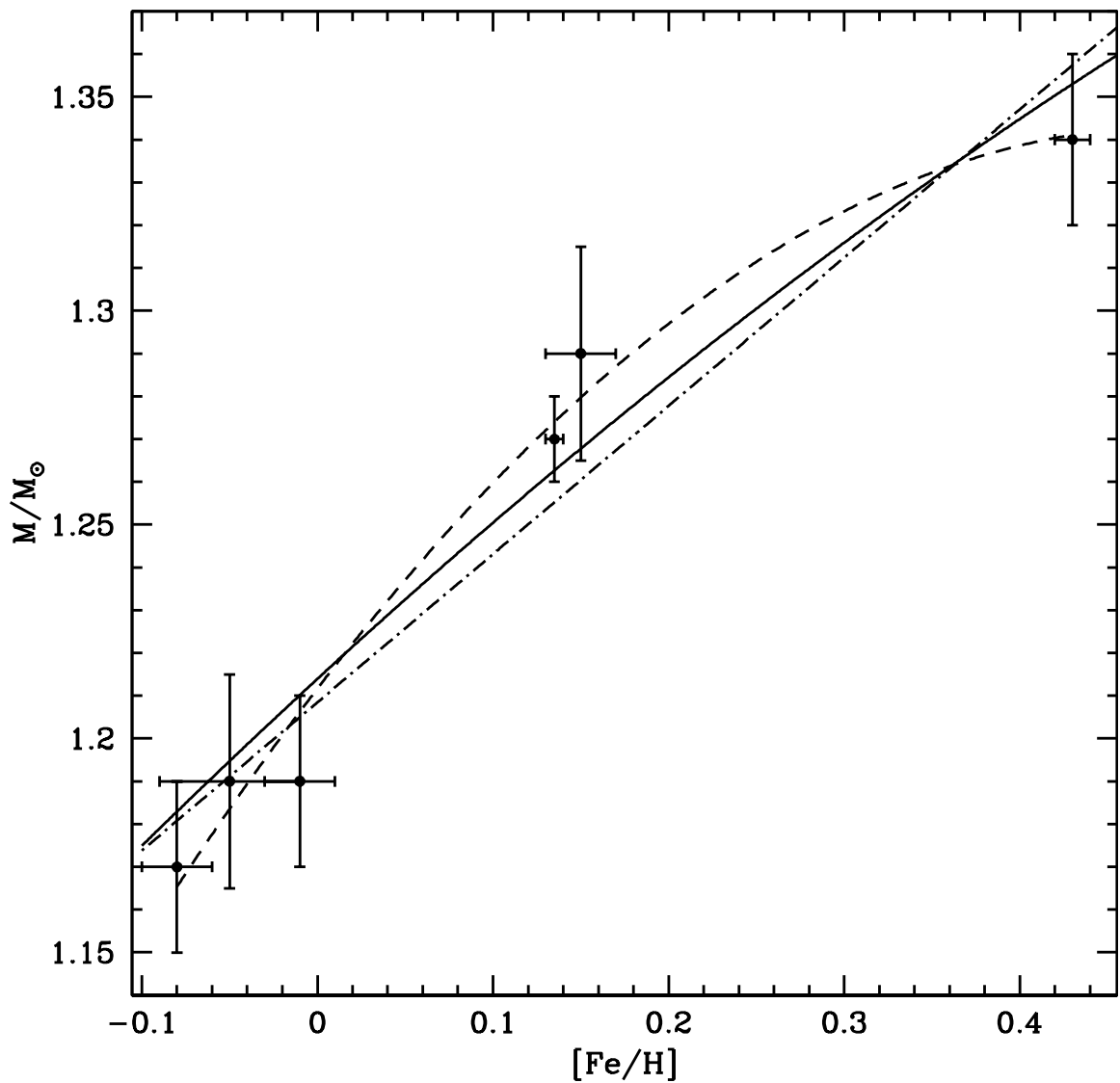}
\end{center}
\caption{The correlation of the gap's cool-side position in mass and the cluster's [Fe/H].  As
observed before (AT09; Chen et al. 2001), the correlation appears nearly linear when looking only at
the lower-metallicity clusters, but with the addition of NGC 6253 the correlation may begin to
turn over at higher metallicity.  The consistency of this correlation across a broad range of
open clusters strengthens the claim that the Li gap's cool side (in mass) does not evolve with age.}
\end{figure}

\begin{center}
\renewcommand{\baselinestretch}{1.1}
{\scriptsize \begin{longtable}[htp]{c c c c c c c c c c c c c c}
\multicolumn{13}{c}%
{{\bfseries \tablename\ \thetable{} - NGC 6253 Stellar Data}} \\
\hline
ID & V & B-V & T$_{eff}$ & V$_r$ & $\sigma($V$_r$) & vsini & A(Li) & $\sigma($A(Li)) & M/M$_\odot$ & 650 Myr T$_{eff}$ & S/N & PM \\
\endfirsthead
\multicolumn{13}{c}%
{{\bfseries \tablename\ \thetable{} - NGC 6253 Stellar Data Continued}} \\
\hline
ID & V & B-V & T$_{eff}$ & V$_r$ & $\sigma($V$_r$) & vsini & A(Li) & $\sigma($A(Li)) & M/M$_\odot$ & 650 Myr T$_{eff}$ & S/N & PM \\
\hline
\hline
\endhead
\hline
\hline
333  & 14.942 & 0.905 & 5790 & -29.82 & 2.28 & 24  & 2.06 & 0.06  &  1.476 & 6563 &  145 & 95 \\
364  & 15.020 & 0.890 & 5837 & -30.72 & 4.03 & 5   & 2.00 & 0.05  &  1.454 & 6521 &  168 & 94 \\
436  & 15.173 & 0.876 & 5881 & -27.85 & 0.38 & 5   & 2.31 & 0.03  &  1.413 & 6439 &  149 & -  \\
505  & 15.368 & 0.848 & 5972 & -29.13 & 0.87 & 5   & 2.12 & 0.06  &  1.363 & 6335 &  131 & 94 \\
565  & 15.478 & 0.872 & 5894 & -28.95 & 0.81 & 5   & 2.40 & 0.04  &  1.337 & 6277 &  110 & 85 \\ 
575  & 15.506 & 0.833 & 6021 & -28.70 & 0.88 & 5   & 2.52 & 0.04  &  1.330 & 6262 &  108 & -  \\
594  & 15.528 & 0.823 & 6055 & -30.90 & 0.96 & 5   & 2.49 & 0.04  &  1.325 & 6250 &  117 & -  \\
628  & 15.594 & 0.874 & 5888 & -28.81 & 0.12 & 5   & 2.55 & 0.03  &  1.309 & 6215 &  113 & 83 \\
645  & 15.626 & 0.864 & 5920 & -30.51 & 0.60 & 5   & 2.38 & 0.05  &  1.302 & 6198 &  93  & -  \\
671  & 15.676 & 0.862 & 5926 & -30.77 & 1.03 & 5   & 2.61 & 0.03  &  1.290 & 6172 &  126 & 79 \\
709  & 15.763 & 0.881 & 5865 & -29.73 & 3.24 & 5   & 2.75 & 0.03  &  1.271 & 6126 &  92  & -  \\
726  & 15.795 & 0.860 & 5933 & -29.21 & 1.54 & 5   & 2.42 & 0.06  &  1.264 & 6108 &  78  & 91 \\
738  & 15.809 & 0.821 & 6061 & -28.85 & 0.13 & 11  & 2.53 & 0.05  &  1.261 & 6101 &  91  & 86 \\ 
758  & 15.844 & 0.878 & 5875 & -29.51 & 0.82 & 5   & 2.31 & 0.04  &  1.253 & 6083 &  132 & 91 \\
770  & 15.868 & 0.872 & 5894 & -31.27 & 1.01 & 5   & 2.39 & 0.06  &  1.248 & 6070 &  72  & 88 \\
777  & 15.888 & 0.841 & 5995 & -28.93 & 0.65 & 5   & 2.33 & 0.08  &  1.243 & 6059 &  71  & 87 \\
874  & 16.042 & 0.867 & 5910 & -29.81 & 0.66 & 5   & 1.97 & 0.11  &  1.211 & 5977 &  99  & 90 \\
932  & 16.121 & 0.866 & 5911 & -29.97 & 1.33 & 5   & 1.64 & 0.15  &  1.195 & 5935 &  74  & -  \\
951  & 16.156 & 0.879 & 5872 & -28.19 & 0.48 & 5   & 2.02 & 0.12  &  1.187 & 5916 &  73  & 3  \\
985  & 16.205 & 0.898 & 5807 & -29.01 & 0.99 & 10  & 1.70 & 0.13  &  1.178 & 5890 &  113 & 87 \\
\hline
\hline
\multicolumn{11}{l}{{3$\sigma$ Upper Limits}}  \\
\hline
193  & 14.464 & 0.886 & 6214 & -28.88 & 0.17 & 5   & 1.70(2) & -  &  1.538 &  6683 &  85  & -  \\
210  & 14.530 & 0.926 & 5719 & -30.29 & 0.54 & 5   & 1.45(1) & -  &  1.539 &  6685 &  65  & -  \\
219  & 14.564 & 0.718 & 6432 & -27.01 & 4.84 & 5   & 1.85(2) & -  &  1.528 &  6664 &  196 & -  \\
224  & 14.573 & 0.977 & 5564 & -32.24 & 0.54 & 5   & 1.10(2) & -  &  1.534 &  6675 &  129 & 92 \\
236  & 14.619 & 0.866 & 5911 & -28.13 & 0.21 & 5   & 1.35(2) & -  &  1.525 &  6658 &  161 & -  \\
250  & 14.662 & 0.847 & 5975 & -30.73 & 4.86 & 5   & 1.40(2) & -  &  1.524 &  6656 &  136 & 92 \\
251  & 14.662 & 1.232 & 4917 & -28.84 & 0.38 & 5   & 0.85(2) & -  &  1.543 &  6693 &  67  & 97 \\
290  & 14.773 & 1.025 & 5426 & -29.63 & 0.12 & 5   & 1.05(2) & -  &  1.536 &  6679 &  113 & 93 \\
314  & 14.891 & 1.066 & 5315 & -29.26 & 2.28 & 5   & 1.00(2) & -  &  1.537 &  6681 &  82  & 96 \\
353  & 15.000 & 1.106 & 5211 & -29.04 & 0.39 & 5   & 1.00(2) & -  &  1.538 &  6683 &  83  & 94 \\
389  & 15.071 & 0.871 & 5895 & -29.31 & 0.62 & 10  & 1.64(1) & -  &  1.440 &  6494 &  146 & 93 \\
401  & 15.099 & 0.854 & 5951 & -30.35 & 2.89 & 5   & 1.35(2) & -  &  1.433 &  6479 &  160 & -  \\
426  & 15.149 & 0.867 & 5908 & -27.36 & 1.63 & 5   & 1.40(2) & -  &  1.419 &  6452 &  136 & 94 \\
451  & 15.223 & 0.828 & 6039 & -29.28 & 1.41 & 5   & 1.40(2) & -  &  1.400 &  6412 &  143 & 85 \\
463  & 15.263 & 0.867 & 5908 & -29.19 & 0.49 & 5   & 1.55(1) & -  &  1.390 &  6391 &  90  & -  \\
503  & 15.365 & 0.883 & 5856 & -29.16 & 0.27 & 5   & 1.40(2) & -  &  1.364 &  6337 &  106 & -  \\
1003 & 16.236 & 0.778 & 5846 & -28.77 & 0.44 & 5   & 1.35(1) & -  &  1.172 &  5873 &  134 & 51 \\
1004 & 16.240 & 0.848 & 5971 & -30.35 & 1.07 & 5   & 2.05(1) & -  &  1.171 &  5871 &  47  & 54 \\
1027 & 16.266 & 0.884 & 5853 & -27.60 & 3.30 & 5   & 1.45(1) & -  &  1.166 &  5857 &  80  & 77 \\
1057 & 16.298 & 0.926 & 5719 & -25.39 & 0.79 & 5   & 1.22(1) & -  &  1.159 &  5840 &  75  & -  \\
1175 & 16.446 & 0.886 & 5846 & -30.23 & 1.03 & 5   & 1.52(1) & -  &  1.131 &  5760 &  75  & 0  \\
1204 & 16.489 & 0.905 & 5785 & -29.94 & 0.33 & 12  & 1.98(1) & -  &  1.123 &  5738 &  39  & 80 \\
\hline
\caption{This table presents information for all of the 42 radial-velocity members.  Star IDs are
the same as in AT10.  Stars with Li detections are listed first and stars with upper limits are
listed at the end.  The listed $\sigma$(A(Li)) for detections is only based on the error from S/N
of the spectra and does not include the systematics discussed earlier.  The two methods discussed
for determining Li abundance upper limits are differentiated by a (1) or a (2) following the
listed upper limit.  The final column gives the PM membership probabilities.  The two
radial-velocity members with less than 5\%\, membership probability are included, but they are not
considered members.}
\end{longtable}}
\end{center}

\begin{center}
\renewcommand{\baselinestretch}{1.1}
{\small \begin{longtable}[htp]{c c c c c c c}
\multicolumn{7}{c}%
{{\bfseries \tablename\ \thetable{} - Cluster Parameters}} \\
\hline
 & NGC 6253 & M67 & Hyades & NGC 3680 & NGC 752 & IC 4651 \\
\hline
\hline
[Fe/H]  &+0.43$\pm$0.01 & -0.009$\pm$0.009  & +0.135$\pm$0.005 & -0.08$\pm$0.02 & -0.05$\pm$0.04 & $\sim$+0.15 \\
age(Gyr) & 3.0$\pm$0.4 & 4.5$\pm$0.5 & $\sim$0.65 & 1.75$\pm$0.1   & 1.45$\pm$0.1   & 1.5$\pm$0.1 \\
E(B-V) & 0.22$\pm$0.04 & 0.041$\pm$0.004 & 0.0 & 0.058 $\pm$0.003 & 0.035 & 0.12 \\
\hline
\caption{This table summarizes the age, metallicity, and reddening for all 6 clusters.  The sources for these
parameters are discussed in the text.}
\end{longtable}}
\end{center}

\end{document}